\documentclass[reprint,aps,prd,superscriptaddress,showpacs,preprintnumbers,amsmath,amssymb,floatfix]{revtex4-1} 

\pdfoutput=1 

\usepackage[mathlines]{lineno}
\usepackage{graphicx,subfigure}
\usepackage{epstopdf}
\usepackage{dcolumn}
\usepackage{bm}
\usepackage{color}
\usepackage{upgreek}
\usepackage{afterpage}
\usepackage{tikz, pgfplots}
\usepackage{bigstrut}
\usepackage[version=3]{mhchem} 
\usepackage[colorlinks=true]{hyperref}

\newcommand{\seq}{Eq.~}

\newcommand{\sfig}{Fig.~}

\newcommand{\stab}{Table~}

\newcommand{\gev}{\nolinebreak GeV/$c^2$}
\newcommand{\keVee}{keV$_{\text{ee}}~$} 
\newcommand{\keVeeNS}{keV$_{\text{ee}}$} 
\newcommand{\keVnrNS}{keV$_{\text{nr}}$} 
\newcommand{\eVee}{eV$_{\text{ee}}~$} 
\newcommand{\eVeeNS}{eV$_{\text{ee}}$} 
\newcommand{\kgd}{kg\,days~} 
\newcommand{\kgdNS}{kg\,days} 

\newcolumntype{1}[1]{D{,}{\pm}{#1}}
\newcolumntype{2}[1]{D{,}{.}{#1}}
\newcolumntype{3}[1]{D{,}{-}{#1}}

\begin{document}


\title{WIMP-Search Results from the Second CDMSlite Run}

\affiliation{Division of Physics, Mathematics, \& Astronomy, California Institute of Technology, Pasadena, CA 91125, USA} 
\affiliation{Institute for Particle Physics Phenomenology, Department of Physics, Durham University, Durham, UK}
\affiliation{Fermi National Accelerator Laboratory, Batavia, IL 60510, USA}
\affiliation{Lawrence Berkeley National Laboratory, Berkeley, CA 94720, USA}
\affiliation{Department of Physics, Massachusetts Institute of Technology, Cambridge, MA 02139, USA}
\affiliation{Department of Physics \& Astronomy, Northwestern University, Evanston, IL 60208-3112, USA}
\affiliation{Pacific Northwest National Laboratory, Richland, WA 99352, USA}
\affiliation{Department of Physics, Queen's University, Kingston, ON K7L 3N6, Canada}
\affiliation{Department of Physics, Santa Clara University, Santa Clara, CA 95053, USA}
\affiliation{SLAC National Accelerator Laboratory/Kavli Institute for Particle Astrophysics and Cosmology, 2575 Sand Hill Road, Menlo Park 94025, CA}
\affiliation{Department of Physics, South Dakota School of Mines and Technology, Rapid City, SD 57701, USA}
\affiliation{Department of Physics, Southern Methodist University, Dallas, TX 75275, USA}
\affiliation{Department of Physics, Stanford University, Stanford, CA 94305, USA}
\affiliation{Department of Physics, Syracuse University, Syracuse, NY 13244, USA}
\affiliation{Department of Physics and Astronomy, and the Mitchell Institute for Fundamental Physics and Astronomy, Texas A\&M University, College Station, TX 77843, USA}
\affiliation{Departamento de F\'{\i}sica Te\'orica and Instituto de F\'{\i}sica Te\'orica UAM/CSIC, Universidad Aut\'onoma de Madrid, 28049 Madrid, Spain}
\affiliation{Department of Physics \& Astronomy, University of British Columbia, Vancouver, BC V6T 1Z1, Canada}
\affiliation{Department of Physics, University of California, Berkeley, CA 94720, USA}
\affiliation{Department of Physics, University of California, Santa Barbara, CA 93106, USA}
\affiliation{Department of Physics, University of Colorado Denver, Denver, CO 80217, USA}
\affiliation{Department of Physics, University of Evansville, Evansville, IN 47722, USA}
\affiliation{Department of Physics, University of Florida, Gainesville, FL 32611, USA}
\affiliation{Department of Physics, University of Illinois at Urbana-Champaign, Urbana, IL 61801, USA}
\affiliation{School of Physics \& Astronomy, University of Minnesota, Minneapolis, MN 55455, USA}
\affiliation{Department of Physics, University of South Dakota, Vermillion, SD 57069, USA}

\author{R.~Agnese} \affiliation{Department of Physics, University of Florida, Gainesville, FL 32611, USA}
\author{A.J.~Anderson} \affiliation{Fermi National Accelerator Laboratory, Batavia, IL 60510, USA}
\author{T.~Aramaki} \affiliation{SLAC National Accelerator Laboratory/Kavli Institute for Particle Astrophysics and Cosmology, 2575 Sand Hill Road, Menlo Park 94025, CA}
\author{M.~Asai} \affiliation{SLAC National Accelerator Laboratory/Kavli Institute for Particle Astrophysics and Cosmology, 2575 Sand Hill Road, Menlo Park 94025, CA}
\author{W.~Baker} \affiliation{Department of Physics and Astronomy, and the Mitchell Institute for Fundamental Physics and Astronomy, Texas A\&M University, College Station, TX 77843, USA}
\author{D.~Balakishiyeva} \affiliation{Department of Physics, University of Florida, Gainesville, FL 32611, USA}
\author{D.~Barker} \affiliation{School of Physics \& Astronomy, University of Minnesota, Minneapolis, MN 55455, USA}
\author{R.~Basu~Thakur~} \affiliation{Fermi National Accelerator Laboratory, Batavia, IL 60510, USA}\affiliation{Department of Physics, University of Illinois at Urbana-Champaign, Urbana, IL 61801, USA}
\author{D.A.~Bauer} \affiliation{Fermi National Accelerator Laboratory, Batavia, IL 60510, USA}
\author{J.~Billard} \affiliation{Department of Physics, Massachusetts Institute of Technology, Cambridge, MA 02139, USA}
\author{A.~Borgland} \affiliation{SLAC National Accelerator Laboratory/Kavli Institute for Particle Astrophysics and Cosmology, 2575 Sand Hill Road, Menlo Park 94025, CA}
\author{M.A.~Bowles} \affiliation{Department of Physics, Syracuse University, Syracuse, NY 13244, USA}
\author{P.L.~Brink} \affiliation{SLAC National Accelerator Laboratory/Kavli Institute for Particle Astrophysics and Cosmology, 2575 Sand Hill Road, Menlo Park 94025, CA}
\author{R.~Bunker} \affiliation{Department of Physics, South Dakota School of Mines and Technology, Rapid City, SD 57701, USA}
\author{B.~Cabrera} \affiliation{Department of Physics, Stanford University, Stanford, CA 94305, USA}
\author{D.O.~Caldwell} \affiliation{Department of Physics, University of California, Santa Barbara, CA 93106, USA}
\author{R.~Calkins} \affiliation{Department of Physics, Southern Methodist University, Dallas, TX 75275, USA}
\author{D.G.~Cerdeno} \affiliation{Institute for Particle Physics Phenomenology, Department of Physics, Durham University, Durham, UK}
\author{H.~Chagani} \affiliation{School of Physics \& Astronomy, University of Minnesota, Minneapolis, MN 55455, USA}
\author{Y.~Chen} \affiliation{Department of Physics, Syracuse University, Syracuse, NY 13244, USA}
\author{J.~Cooley} \affiliation{Department of Physics, Southern Methodist University, Dallas, TX 75275, USA}
\author{B.~Cornell} \affiliation{Division of Physics, Mathematics, \& Astronomy, California Institute of Technology, Pasadena, CA 91125, USA}
\author{P.~Cushman} \affiliation{School of Physics \& Astronomy, University of Minnesota, Minneapolis, MN 55455, USA}
\author{M.~Daal} \affiliation{Department of Physics, University of California, Berkeley, CA 94720, USA}
\author{P.C.F.~Di~Stefano} \affiliation{Department of Physics, Queen's University, Kingston, ON K7L 3N6, Canada}
\author{T.~Doughty} \affiliation{Department of Physics, University of California, Berkeley, CA 94720, USA}
\author{L.~Esteban} \affiliation{Departamento de F\'{\i}sica Te\'orica and Instituto de F\'{\i}sica Te\'orica UAM/CSIC, Universidad Aut\'onoma de Madrid, 28049 Madrid, Spain}
\author{S.~Fallows} \affiliation{School of Physics \& Astronomy, University of Minnesota, Minneapolis, MN 55455, USA}
\author{E.~Figueroa-Feliciano} \affiliation{Department of Physics \& Astronomy, Northwestern University, Evanston, IL 60208-3112, USA}
\author{M.~Ghaith} \affiliation{Department of Physics, Queen's University, Kingston, ON K7L 3N6, Canada}
\author{G.L.~Godfrey} \affiliation{SLAC National Accelerator Laboratory/Kavli Institute for Particle Astrophysics and Cosmology, 2575 Sand Hill Road, Menlo Park 94025, CA}
\author{S.R.~Golwala} \affiliation{Division of Physics, Mathematics, \& Astronomy, California Institute of Technology, Pasadena, CA 91125, USA}
\author{J.~Hall} \affiliation{Pacific Northwest National Laboratory, Richland, WA 99352, USA}
\author{H.R.~Harris} \affiliation{Department of Physics and Astronomy, and the Mitchell Institute for Fundamental Physics and Astronomy, Texas A\&M University, College Station, TX 77843, USA}
\author{T.~Hofer} \affiliation{School of Physics \& Astronomy, University of Minnesota, Minneapolis, MN 55455, USA}
\author{D.~Holmgren} \affiliation{Fermi National Accelerator Laboratory, Batavia, IL 60510, USA}
\author{L.~Hsu} \affiliation{Fermi National Accelerator Laboratory, Batavia, IL 60510, USA}
\author{M.E.~Huber} \affiliation{Department of Physics, University of Colorado Denver, Denver, CO 80217, USA}
\author{D.~Jardin} \affiliation{Department of Physics, Southern Methodist University, Dallas, TX 75275, USA}
\author{A.~Jastram} \affiliation{Department of Physics and Astronomy, and the Mitchell Institute for Fundamental Physics and Astronomy, Texas A\&M University, College Station, TX 77843, USA}
\author{O.~Kamaev} \affiliation{Department of Physics, Queen's University, Kingston, ON K7L 3N6, Canada}
\author{B.~Kara} \affiliation{Department of Physics, Southern Methodist University, Dallas, TX 75275, USA}
\author{M.H.~Kelsey} \affiliation{SLAC National Accelerator Laboratory/Kavli Institute for Particle Astrophysics and Cosmology, 2575 Sand Hill Road, Menlo Park 94025, CA}
\author{A.~Kennedy} \affiliation{School of Physics \& Astronomy, University of Minnesota, Minneapolis, MN 55455, USA}
\author{A.~Leder} \affiliation{Department of Physics, Massachusetts Institute of Technology, Cambridge, MA 02139, USA}
\author{B.~Loer} \affiliation{Fermi National Accelerator Laboratory, Batavia, IL 60510, USA}
\author{E.~Lopez~Asamar} \affiliation{Departamento de F\'{\i}sica Te\'orica and Instituto de F\'{\i}sica Te\'orica UAM/CSIC, Universidad Aut\'onoma de Madrid, 28049 Madrid, Spain}
\author{P.~Lukens} \affiliation{Fermi National Accelerator Laboratory, Batavia, IL 60510, USA}
\author{R.~Mahapatra} \affiliation{Department of Physics and Astronomy, and the Mitchell Institute for Fundamental Physics and Astronomy, Texas A\&M University, College Station, TX 77843, USA}
\author{V.~Mandic} \affiliation{School of Physics \& Astronomy, University of Minnesota, Minneapolis, MN 55455, USA}
\author{N.~Mast} \affiliation{School of Physics \& Astronomy, University of Minnesota, Minneapolis, MN 55455, USA}
\author{N.~Mirabolfathi} \affiliation{Department of Physics, University of California, Berkeley, CA 94720, USA}
\author{R.A.~Moffatt} \affiliation{Department of Physics, Stanford University, Stanford, CA 94305, USA}
\author{J.D.~Morales~Mendoza} \affiliation{Department of Physics and Astronomy, and the Mitchell Institute for Fundamental Physics and Astronomy, Texas A\&M University, College Station, TX 77843, USA}
\author{S.M.~Oser} \affiliation{Department of Physics \& Astronomy, University of British Columbia, Vancouver, BC V6T 1Z1, Canada}
\author{K.~Page} \affiliation{Department of Physics, Queen's University, Kingston, ON K7L 3N6, Canada}
\author{W.A.~Page} \affiliation{Department of Physics \& Astronomy, University of British Columbia, Vancouver, BC V6T 1Z1, Canada}
\author{R.~Partridge} \affiliation{SLAC National Accelerator Laboratory/Kavli Institute for Particle Astrophysics and Cosmology, 2575 Sand Hill Road, Menlo Park 94025, CA}
\author{M.~Pepin} \email{Corresponding author: pepin@physics.umn.edu} \affiliation{School of Physics \& Astronomy, University of Minnesota, Minneapolis, MN 55455, USA}
\author{A.~Phipps} \affiliation{Department of Physics, University of California, Berkeley, CA 94720, USA}
\author{K.~Prasad} \affiliation{Department of Physics and Astronomy, and the Mitchell Institute for Fundamental Physics and Astronomy, Texas A\&M University, College Station, TX 77843, USA}
\author{M.~Pyle} \affiliation{Department of Physics, University of California, Berkeley, CA 94720, USA}
\author{H.~Qiu} \affiliation{Department of Physics, Southern Methodist University, Dallas, TX 75275, USA}
\author{W.~Rau} \affiliation{Department of Physics, Queen's University, Kingston, ON K7L 3N6, Canada}
\author{P.~Redl} \affiliation{Department of Physics, Stanford University, Stanford, CA 94305, USA}
\author{A.~Reisetter} \affiliation{Department of Physics, University of Evansville, Evansville, IN 47722, USA}
\author{Y.~Ricci} \affiliation{Department of Physics, Queen's University, Kingston, ON K7L 3N6, Canada}
\author{A.~Roberts} \affiliation{Department of Physics, University of South Dakota, Vermillion, SD 57069, USA}
\author{H.E.~Rogers} \affiliation{School of Physics \& Astronomy, University of Minnesota, Minneapolis, MN 55455, USA}
\author{T.~Saab} \affiliation{Department of Physics, University of Florida, Gainesville, FL 32611, USA}
\author{B.~Sadoulet} \affiliation{Department of Physics, University of California, Berkeley, CA 94720, USA}\affiliation{Lawrence Berkeley National Laboratory, Berkeley, CA 94720, USA}
\author{J.~Sander} \affiliation{Department of Physics, University of South Dakota, Vermillion, SD 57069, USA}
\author{K.~Schneck} \affiliation{SLAC National Accelerator Laboratory/Kavli Institute for Particle Astrophysics and Cosmology, 2575 Sand Hill Road, Menlo Park 94025, CA}
\author{R.W.~Schnee} \affiliation{Department of Physics, South Dakota School of Mines and Technology, Rapid City, SD 57701, USA}
\author{S.~Scorza} \affiliation{Department of Physics, Southern Methodist University, Dallas, TX 75275, USA}
\author{B.~Serfass} \affiliation{Department of Physics, University of California, Berkeley, CA 94720, USA}
\author{B.~Shank} \affiliation{Department of Physics, Stanford University, Stanford, CA 94305, USA}
\author{D.~Speller} \affiliation{Department of Physics, University of California, Berkeley, CA 94720, USA}
\author{D.~Toback} \affiliation{Department of Physics and Astronomy, and the Mitchell Institute for Fundamental Physics and Astronomy, Texas A\&M University, College Station, TX 77843, USA}
\author{R.~Underwood} \affiliation{Department of Physics, Queen's University, Kingston, ON K7L 3N6, Canada}
\author{S.~Upadhyayula} \affiliation{Department of Physics and Astronomy, and the Mitchell Institute for Fundamental Physics and Astronomy, Texas A\&M University, College Station, TX 77843, USA}
\author{A.N.~Villano} \affiliation{School of Physics \& Astronomy, University of Minnesota, Minneapolis, MN 55455, USA}
\author{B.~Welliver} \affiliation{Department of Physics, University of Florida, Gainesville, FL 32611, USA}
\author{J.S.~Wilson} \affiliation{Department of Physics and Astronomy, and the Mitchell Institute for Fundamental Physics and Astronomy, Texas A\&M University, College Station, TX 77843, USA}
\author{D.H.~Wright} \affiliation{SLAC National Accelerator Laboratory/Kavli Institute for Particle Astrophysics and Cosmology, 2575 Sand Hill Road, Menlo Park 94025, CA}
\author{S.~Yellin} \affiliation{Department of Physics, Stanford University, Stanford, CA 94305, USA}
\author{J.J.~Yen} \affiliation{Department of Physics, Stanford University, Stanford, CA 94305, USA}
\author{B.A.~Young} \affiliation{Department of Physics, Santa Clara University, Santa Clara, CA 95053, USA}
\author{J.~Zhang} \affiliation{School of Physics \& Astronomy, University of Minnesota, Minneapolis, MN 55455, USA}

\smallskip
\date{\today}

\collaboration{SuperCDMS Collaboration}

\noaffiliation


\smallskip

\begin{abstract}
The CDMS low ionization threshold experiment (CDMSlite) uses cryogenic germanium detectors operated at a relatively high bias voltage to amplify the phonon signal in the search for weakly interacting massive particles (WIMPs). Results are presented from the second CDMSlite run with an exposure of 70~\kgdNS, which reached an energy threshold for electron recoils as low as 56~eV. A fiducialization cut reduces backgrounds below those previously reported by CDMSlite. New parameter space for the WIMP-nucleon spin-independent cross section is excluded for WIMP masses between 1.6 and 5.5~\gev.
\end{abstract}


\maketitle

%
%
%
%
%
%
%
%
%
%
%
%

Cosmological and astrophysical measurements indicate that over one quarter of the energy density of the universe consists of nonbaryonic and nonluminous matter~\cite{Olive2014,Ade2014}.   Weakly interacting massive particles (WIMPs) produced in the Big Bang are a compelling class of candidate particles for this dark matter~\cite{Jungman1996}.   
Recent accelerator~\cite{Abercrombie2015} and direct-detection results~\cite{Akerib2014} constrain the simplest supersymmetric models, sparking interest in alternative theories, including theories with low-mass dark matter~\cite{PhysRevD.86.056009,Kusenko20091,PhysRevD.78.115012,PhysRevD.75.115017,PhysRevLett.101.231301,Boehm2004219}.
It has been suggested that, because the energy densities are similar for baryonic and dark matter, the relic density may be generated by an asymmetry related to the baryon asymmetry~\cite{Kaplan1992,Kaplan2009,Falkowski2011,Volkas2013,Zurek2014}.  In this case,  the number densities of the two are also related, suggesting searches for particles with masses of a few~\gev. Hints of WIMP signals near detector thresholds ~\cite{Agnese2013,Angloher2012,Aalseth2013,Savage2009,Bernabei2010}, and an excess of gamma-ray emission from the Galactic Center~\cite{Hooper2011}, have also contributed to interest in the low-mass WIMP region~\cite{Agnese2014,Agnese2014a,Armengaud2012,Akerib2014,Angloher2014,Angloher2015,Barreto2012,Amole2015,Yue2014,Li2013,Angle2011,*Angle2013}.

WIMPs may create keV-scale recoils in laboratory detectors by elastically scattering off target nuclei~\cite{Goodman1985}.
WIMPs significantly below the target nuclei mass are an experimental challenge because they deposit small recoil energies, making it harder to distinguish WIMP signals from background and electronic noise. At these energies, techniques used to discriminate between nuclear recoils (NRs) and electron recoils (ERs) often diminish in effectiveness.
The CDMS low ionization threshold experiment (CDMSlite)~\cite{Agnese2014} exchanges NR discrimination for a lower threshold. This paper describes data taken with one SuperCDMS iZIP detector~\cite{Agnese2013a} operated in CDMSlite mode at the Soudan Underground Laboratory~\cite{Akerib2005}.

CDMSlite employs the Neganov-Luke effect~\cite{Luke1988,Neganov1985,Wang2010}, which amplifies phonon signals when electric charges drift through a material~\cite{Akerib2004,Isaila2012,Spooner1992}. Work done by the electric field on the electron-hole pairs is fully converted into phonons along the drift path of the charges.  The total measured phonon signal $E_t$, including the energy $E_r$ from the initial particle interaction, is
\begin{equation}
	\label{eq:lukeAmp}
	E_t = E_r + N_\text{eh}eV_b~,
\end{equation}
where $e$ is the elementary charge, $V_b$ is the bias voltage, and $N_\text{eh}$ is the number of electron-hole pairs created in the interaction.  The average energy required to create an electron-hole pair for an electron recoil in germanium is $\epsilon_{\gamma}=3$~eV/pair, which gives $N_\text{eh}=E_r/\epsilon_{\gamma}$.  When $V_b$ is large compared to $\epsilon_{\gamma}$, the Neganov-Luke phonons dominate the measured signal and allow for lower thresholds to be reached.

Nuclear recoils produce electron-hole pairs less efficiently, so an ionization yield $Y{\left(E_r\right)}$ that depends on energy and interaction type is defined through $N_\text{eh} = Y{\left(E_r\right)}E_r/\epsilon_{\gamma}$, where $Y\equiv1$ for electron recoils. The total energy can be expressed as
\begin{equation}
	\label{eq:lukeAmp2}
	E_t = E_r\left(1+Y{\left(E_r\right)}\frac{eV_b}{\epsilon_{\gamma}}\right).
\end{equation}
The energy scale thus also depends on the interaction type.  The detector is calibrated using ERs and the resulting energy scale is labeled as \keVeeNS.  The spectrum is then converted to nuclear recoil equivalent energy (labeled as \keVnrNS) by comparing \seq\ref{eq:lukeAmp2} for ERs and NRs and solving for $E_{\text{nr}}$:
\begin{equation}
	\label{eq:ee2nr}
	E_{\text{nr}} = E_{\text{ee}}\left(\frac{1+eV_b/\epsilon_{\gamma}}{1+Y{\left(E_{\text{nr}}\right)}eV_b/\epsilon_{\gamma}}\right),
\end{equation}
where $Y{\left(E_{\text{nr}}\right)}$ is the ionization yield for NRs.

During a first short data run of CDMSlite, this mode of operation was proven viable and provided leading sensitivity to WIMPs with masses between 3 and 6~\gev~\cite{Agnese2014}.  The data presented here are from the second CDMSlite run using the same detector, performed from February to November 2014 and taken in three hour long data series (to maintain the neutralization of the detector~\cite{Akerib2005}).  A voltage bias of $-70$~V was applied to one side of the detector with the other side at 0~V.  The electronics setup followed that of the first CDMSlite run with phonon sensors instrumented only on the grounded side of the detector~\cite{Agnese2014}.  The single CDMSlite detector was still part of the full SuperCDMS detector array where all detectors are read-out with every trigger.  Several hardware and operational improvements were implemented for this run~\cite{BasuThakur2014}.  Fluctuations in bias voltage were reduced by cleaning the high-voltage biasing-electronics board, sealing it against humidity, and placing it in a dry nitrogen atmosphere.  The large detector leakage current observed at the start of each data series in the first run was decreased by ``prebiasing'' the detector at $-80$~V for 10 min prior to each series.  Cryocooler-induced microphonic noise that limited the threshold of the first CDMSlite run was better rejected by the installation of vibration sensors near the connection of the cryocooler to the SuperCDMS cryostat.

The cryocooler degraded during the run, causing the induced noise to dominate the trigger rate.  The experiment was warmed to room temperature in July to allow for the routine replacement of the cryocooler cold head.  The run resumed in September and, due to the maintenance reducing the microphonic noise rate, a lower energy threshold was achieved.  The run is thus naturally split into two periods: February--July (period 1) and September--November (period 2).

Time intervals with exceptionally high trigger rates were removed from the WIMP-search exposure.  Events with elevated prepulse noise were conservatively removed because the integrity of the detector cannot be guaranteed during the time since the preceding event.  Glitches, defined as pulses with uncharacteristically sharp rise and fall times, were observed and removed in three categories: electronic glitch events that cause simultaneous triggers in multiple detectors, glitches in the outer charge channel of the CDMSlite detector, which do not cause triggers themselves but can be coincident with a phonon trigger, and glitches that mimic standard events in all but pulse shape. Events coincident with the NuMI neutrino beam~\cite{Michael2008}, and events whose NuMI timing information (for determining coincidence) was unavailable, were removed. The combination of the above cuts, except the third class of glitches (see below), reduced the live time from 132.23 to 115.59~d: 97.81~d (110.28~raw) in period 1 and 17.78~d (21.95~raw) in period 2.  Events with energy deposited in multiple detectors or coincident with the muon veto were also removed, with a combined efficiency of $97.21\pm0.01$\,\% for the detection of dark matter.

Neutron calibrations were performed three times over the course of the run by exposing the detectors to a \ce{^{252}Cf} source.  Neutron capture on \ce{^{70}Ge} creates \ce{^{71}Ge}, which decays via electron capture with a half-life of 11.43~d~\cite{Hampel1985}.  X rays and Auger electrons are emitted with a total energy corresponding to the \ce{^{71}Ga} electron binding energy of the shell from which the electron is captured.  The $K$-, $L$-, and $M$-shell binding energies are 10.37, 1.30, and 0.16~keV, respectively~\cite{Bearden1967}, with the latter two in the energy region of interest for this analysis.

The ER energy scale was calibrated using the K-capture line from \ce{^{71}Ge}.  Drifts in detector bias and cryostat base temperature caused this line to vary by 5--10\% over time.  After correcting for those effects, two small (${\sim}2.5$\,\% total) residual shifts of unknown origin were observed and corrected.

The trigger efficiency was calculated by measuring, as a function of the reconstructed energy in the CDMSlite detector, the fraction of events triggering another detector that also trigger the CDMSlite detector.  The data used for this were a subset of the \ce{^{252}Cf} calibration events, with very strict cuts removing events that could cause triggers due to noise or cross talk in the CDMSlite detector in coincidence with the other iZIP detectors.
In the two periods, 50\,\% trigger efficiency was reached at $75^{+4}_{-5}$ and $56^{+6}_{-4}$~\eVeeNS, respectively, except for the 4.22~days immediately prior to the midrun warm-up.  During these few days, the hardware trigger threshold was increased slightly to reduce noise-induced triggers (see below).

\begin{figure}
	\centering
		\includegraphics[width=\columnwidth]{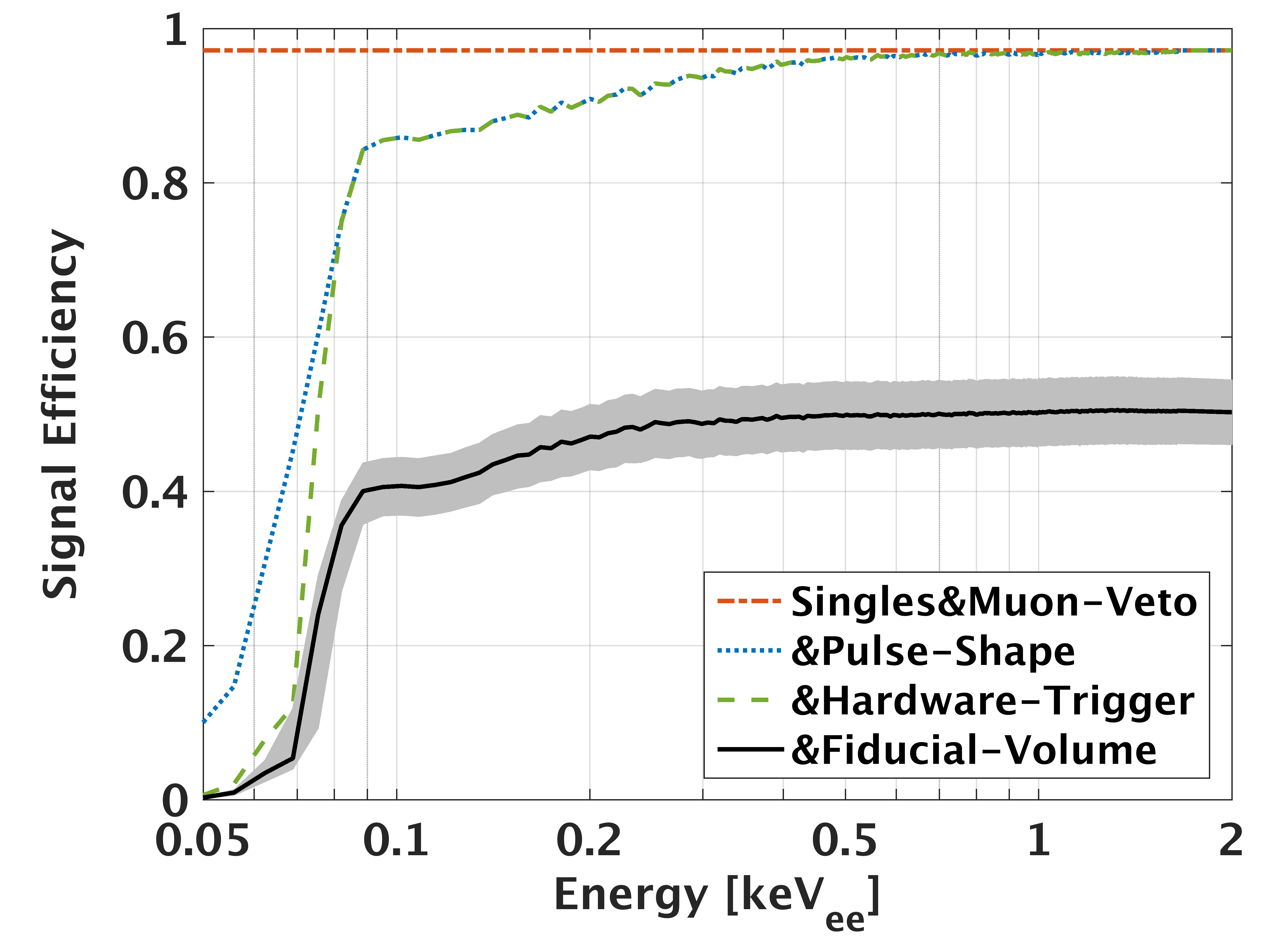}
		\caption{(color online) Binned total signal efficiency after sequential application of selection criteria: single-scatter and muon-veto (orange dashed-dot curve), pulse-shape (blue dotted curve), hardware-trigger (green dashed curve), and radial fiducial-volume (black solid curve) criteria.  The gray band around the final curve shows the combined uncertainty on the overall efficiency at $1\sigma$ confidence.}
		\label{fig:tieredEff}
\end{figure}

Pulse shape was used to distinguish signal events from noise events, to make fiducial cuts, and to correct energy calibrations.  All events were fit to three different templates corresponding to the standard signal event, electronic glitches with quick rise and fall times, and  cryocooler-induced low-frequency noise (LF noise).  The differences between goodness-of-fit quantities for the templates were used to identify glitches and LF noise.  As the cryocooler degraded during period 1, the rate of LF noise greatly increased.  A metric based upon the noise profile across the cryocooler's 830~ms cycle was used to identify periods of calendar time with high or low cryocooler-induced LF noise.  The pulse-shape discrimination cut was then set to be tighter during periods of more cryocooler noise, with the goal of minimizing leakage at the expense of efficiency.  The efficiency of all three pulse-shape criteria was determined by a Monte Carlo pulse simulation that combines experimental noise (taken throughout the entire run) with the standard pulse template, scaled to the desired energies.  Figure~\ref{fig:tieredEff} gives the result for all three pulse-shape cuts, with the dominant loss in efficiency caused by the LF noise cut. Because the energy-estimating algorithm assigns energies of ${\sim}80$~\eVee to LF noise, the goodness-of-fit separation between LF noise and good pulses becomes less distinct near this point.  This requires a harder cut at this energy, leading to a sharp drop in efficiency.  A small systematic uncertainty corresponding to the variation of pulse shape with energy and position was estimated by measuring the efficiency for a range of pulse-shape templates.

\begin{figure}
	\centering
		\includegraphics[width=\columnwidth]{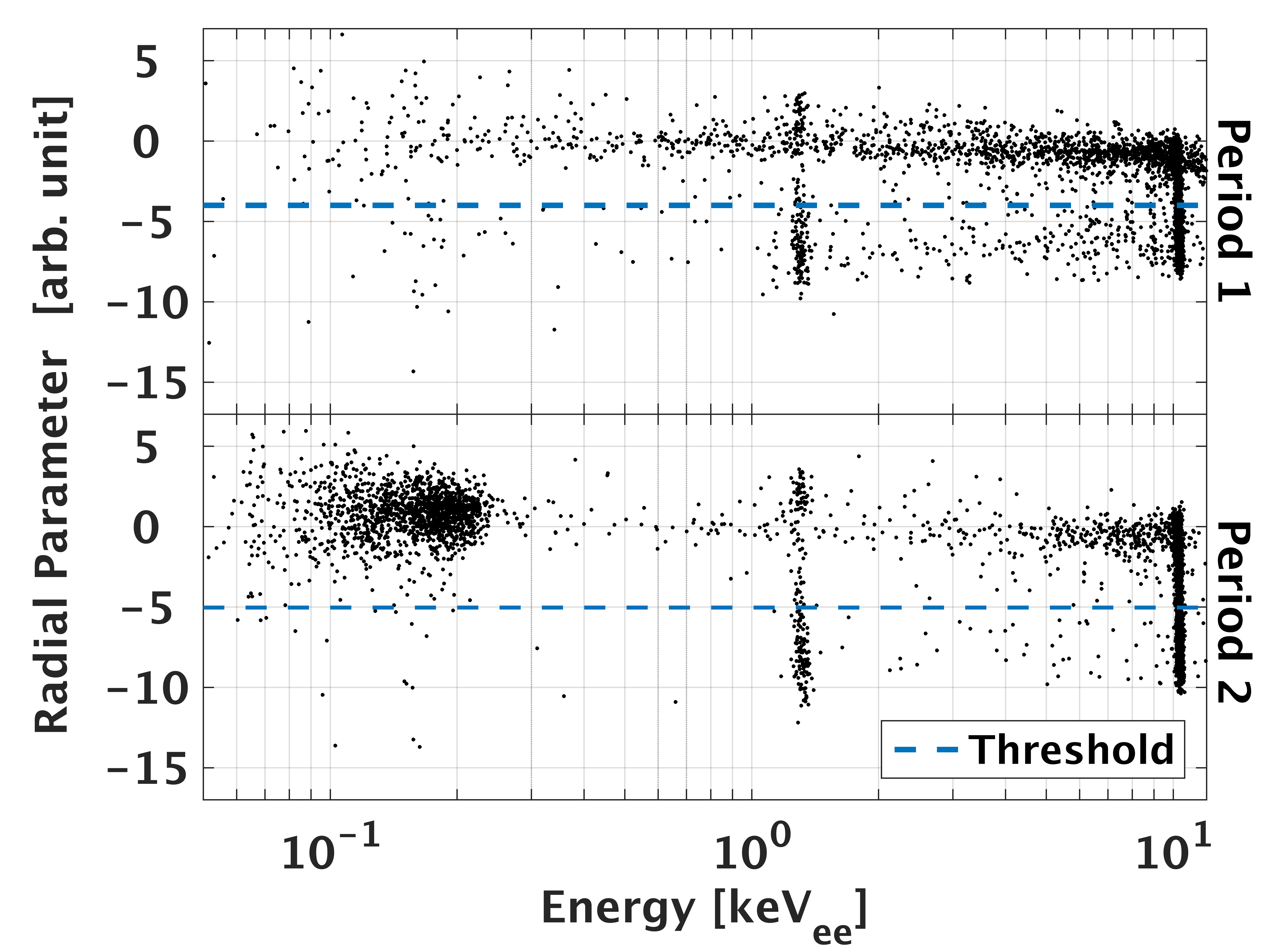}
		\caption{Radial parameter as a function of energy for the first (top) and second (bottom) periods. The dashed lines indicate the radial cut.  The densely populated band at larger values corresponds to events near the edge of the detector.  The vertical clusters are the \ce{^{71}Ge} capture peaks.  The clear separation between outer and inner events decreases at high energy due to signal saturation in the outer phonon channel.  The slight downward shift in the distribution after the maintenance period (caused by a small change in the operating point of the phonon sensors) together with the appearance of an unexplained localized background below ${\sim}$250~\eVee near the edge of the detector motivated a tighter radial cut in period 2.}
		\label{fig:radPlane}
\end{figure}

A valid pulse shape has two components: a fast one whose amplitude depends on the position of the scattering event and a slow one that carries the primary energy information.  For this analysis, a new algorithm was introduced that fits pulses from each phonon channel with a linear combination of fast and slow template pulses, allowing the position and energy information to be separated.  The position information was used, along with the segmentation of the phonon sensor into one outer and three inner segments, to construct a radial parameter by comparing the start time and the amplitude of the fast component of the outer and inner sensors.
This radial parameter is shown as a function of energy in \sfig\ref{fig:radPlane}, where the densely populated band at higher parameter values corresponds to events in the outer part of the detector. A nonuniform field in this region draws charge carriers to the sidewall of the detector, preventing them from traversing the full potential.  This produces a reduced Neganov-Luke amplification and distorts the energy spectrum.

 The radial cut removes more than 90\,\% of these events, along with a small contribution of low-energy surface events originating on the detector housing, down to low energies while maintaining a reasonable fraction of the exposure for inner events. The few remaining reduced-energy events contribute to the background at lower energy.
 In period 2, a cluster of background events appeared below ${\sim}$250~\eVee and was located in the outer part of the detector near one of the connectors. This, together with differences in the operating conditions between the two periods, motivated a tighter cut in period 2.  The fiducial-volume cut significantly reduced the background rate compared to the first CDMSlite run.

The acceptance for the radial fiducial cut was determined using the \ce{^{71}Ge} electron-capture events, which sample a uniform spatial distribution in the detector. These events can be divided into three categories: those degraded in energy, those with full energy that fail the fiducial cut, and those with full energy that pass the fiducial cut. The fraction of events in the first category is given by the electric-field geometry and is energy independent.  To measure this effect, the radius-energy plane was divided into sections and a likelihood-based Monte Carlo simulation was applied to each section independently to determine the contribution of two components: a time-independent background and a contribution from the \ce{^{71}Ge} activation lines exponentially decaying in time.  The known ratio of $L$- to $K$-capture rates was used to separate the $L$- and $K$-capture contributions. The fraction of events with a full Neganov-Luke phonon signal was determined to be ${\sim}$86\,\%.

Next, the fraction of events with full phonon signal removed by the radial cut was computed at the capture-peak locations as the number of events passing the cut criterion divided by all peak events after background subtraction.  The background in the inner part of the detector is negligible compared to the peak rate; in the outer part the background was calculated from the observed event rates above and below the peak.  To measure the efficiency at lower energies, a pulse-simulation method was implemented.  All events from the $L$-capture peak (chosen to avoid observed signal saturation in the outer phonon channel above ${\sim}2$~\keVeeNS) were used to generate nearly noise-free pulses using the extracted composition of the fast and slow templates.  These noise-free pulses were then scaled to the desired energy before adding measured noise. This sample of artificial raw events was analyzed in the same manner as the real raw data. The efficiency was measured using the fraction of artificial events passing the radial cut, taking into account the background contribution in the original event sample.  The combined fiducial-volume efficiency was calculated to be ${\sim}$50\,\% with a mild energy dependence as shown in \sfig\ref{fig:tieredEff}.

\begin{figure}
	\centering
	\includegraphics[width=\columnwidth]{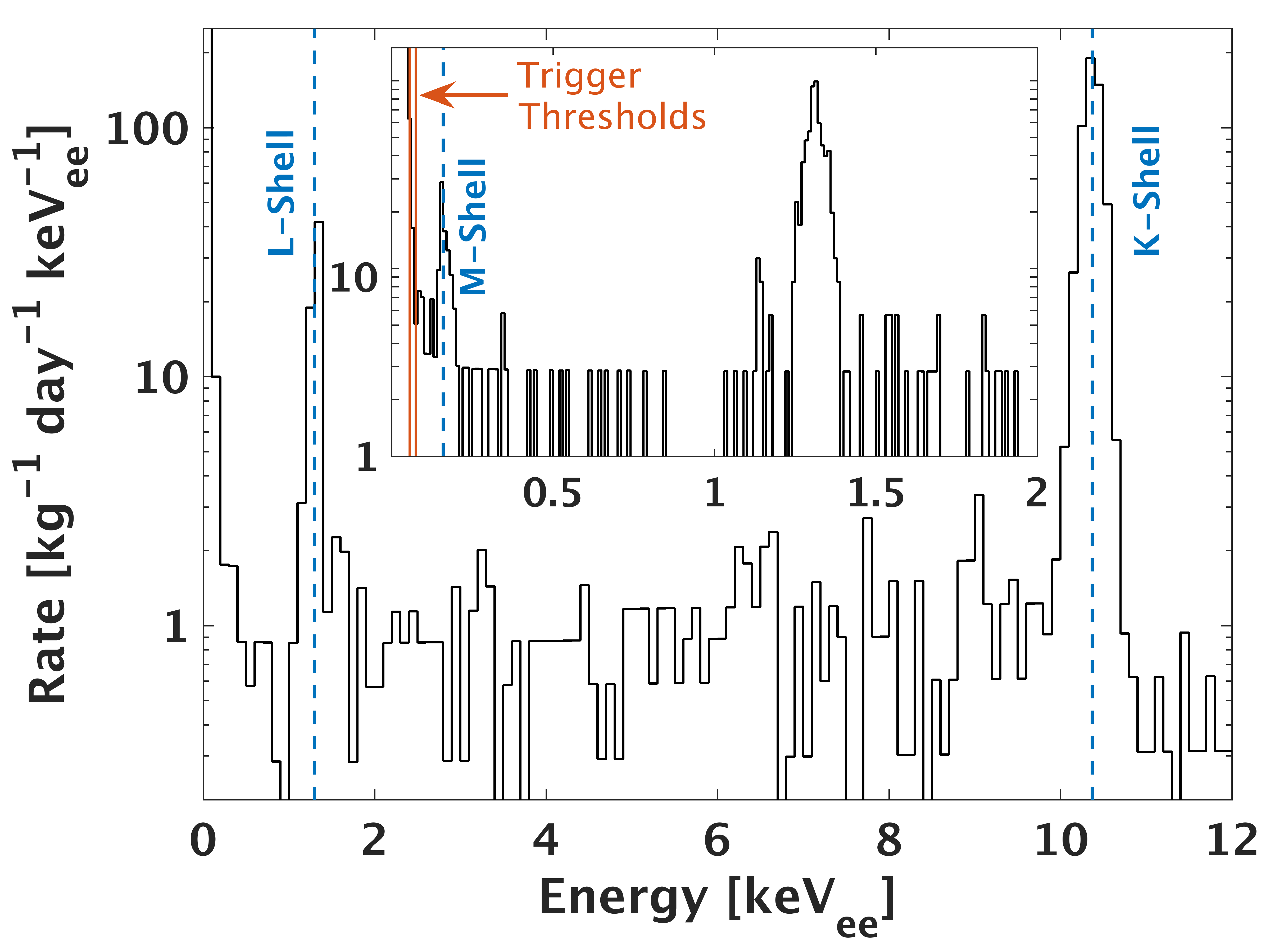}
	\caption{Spectrum of events passing all selection criteria, corrected for all efficiencies except the trigger efficiency.  Dashed lines indicate the prominent features of the \ce{^{71}Ge} electron-capture decay with peaks at 10.37~keV ($K$-shell), 1.30~keV ($L$-shell) and 0.16~keV ($M$-shell).  Inset: Englargement of the lowest energies that determine the low-mass WIMP sensitivity, including the $L$- and $M$-shell activation peaks.  Solid vertical lines show the 50\,\% trigger-efficiency points for the two periods.}
	\label{fig:fullSpec}
\end{figure}

The final spectrum after application of all selection criteria and correcting for all efficiencies (except the trigger efficiency) is shown in \sfig\ref{fig:fullSpec}.  The main features are the \ce{^{71}Ge} electron-capture peaks at 10.37, 1.30, and 0.16~keV.  Hints of other peaks can be seen on top of a smooth background from Compton scattering of higher-energy gamma rays.  The observed ratio of the rates for the $M$- and $L$-capture peaks is $0.16\pm0.03$, compared to an expected $0.17$~\cite{Schonfeld1998}.  Other numerical characterizations of the primary components of the spectrum are listed in \stab\ref{tab:specChar}.

\begin{table}
	\centering
	\begin{tabular}{@{}2{2.2}1{4.4}||3{4.4}c@{}}
		\hline\hline
		\multicolumn{1}{@{}c}{Energy} & \multicolumn{1}{c||}{Resolution} & \multicolumn{1}{c}{Range} & Average rate \bigstrut[t]\\
		\multicolumn{1}{@{}c}{$\left[\text{\keVeeNS}\right]$} & \multicolumn{1}{c||}{$\left[\sigma/\mu,\%\right]$} & \multicolumn{1}{c}{$\left[\text{\keVeeNS}\right]$} & $\left[\text{\keVeeNS\,kg\,day}\right]^{-1}$ \bigstrut[b]\\ \hline
		&          & 0.056,0.14 & $16.33^{+8.18}_{-7.97}$ \bigstrut[t]\\
		0,16  & 11.4,2.8 & 0.2,1.2 & $1.09{\pm}0.18$ \\
		1,30  & 2.36,0.15 & 1.4,10  & $1.00{\pm}0.06$ \\
		10,37 & 0.974,0.009 & 11,20   & $0.30{\pm}0.03$ \bigstrut[b]\\
		\hline\hline
	\end{tabular}
	\caption{\textit{Left}: resolution of the \ce{^{71}Ge} capture peaks.  \textit{Right}: average rate between the peaks, after application of all selection criteria, corrected for efficiency.  The difference in rates above and below the $K$-capture peak can be attributed to unresolved peaks due to cosmogenic backgrounds and the higher rate below the $M$-capture peak can be attributed to more background leakage at lower energies in period 2.}
	\label{tab:specChar}
\end{table}

In CDMSlite mode, the ionization yield cannot be measured on an event-by-event basis, necessitating a model.  The most common model in the field is that of Lindhard~\cite{Lindhard1963,*Lindhard1963a,*Lindhard1968}:
\begin{equation}
	\label{eq:lindhard}
	Y{\left(E_{nr}\right)} = \frac{k\cdot g\left(\varepsilon\right)}{1+k\cdot g\left(\varepsilon\right)}~,
\end{equation}
where $g\left(\varepsilon\right)=3\varepsilon^{0.15}+0.7\varepsilon^{0.6}+\varepsilon$, $\varepsilon=11.5E_{\text{nr}}{\left(\text{keV}\right)}Z^{-7/3}$, and $Z$ is the atomic number of the material.  For germanium, $k=0.157$, but the model is somewhat uncertain for low recoil energies.  This uncertainty was accounted for by varying $k$ uniformly between $k=0.1$ and 0.2, which encompasses a majority of the experimentally observed data~\cite{Barker2012,*Jones1975,*Barbeau2007}.

\begin{figure}[t]
	\centering
		\includegraphics[width=\columnwidth]{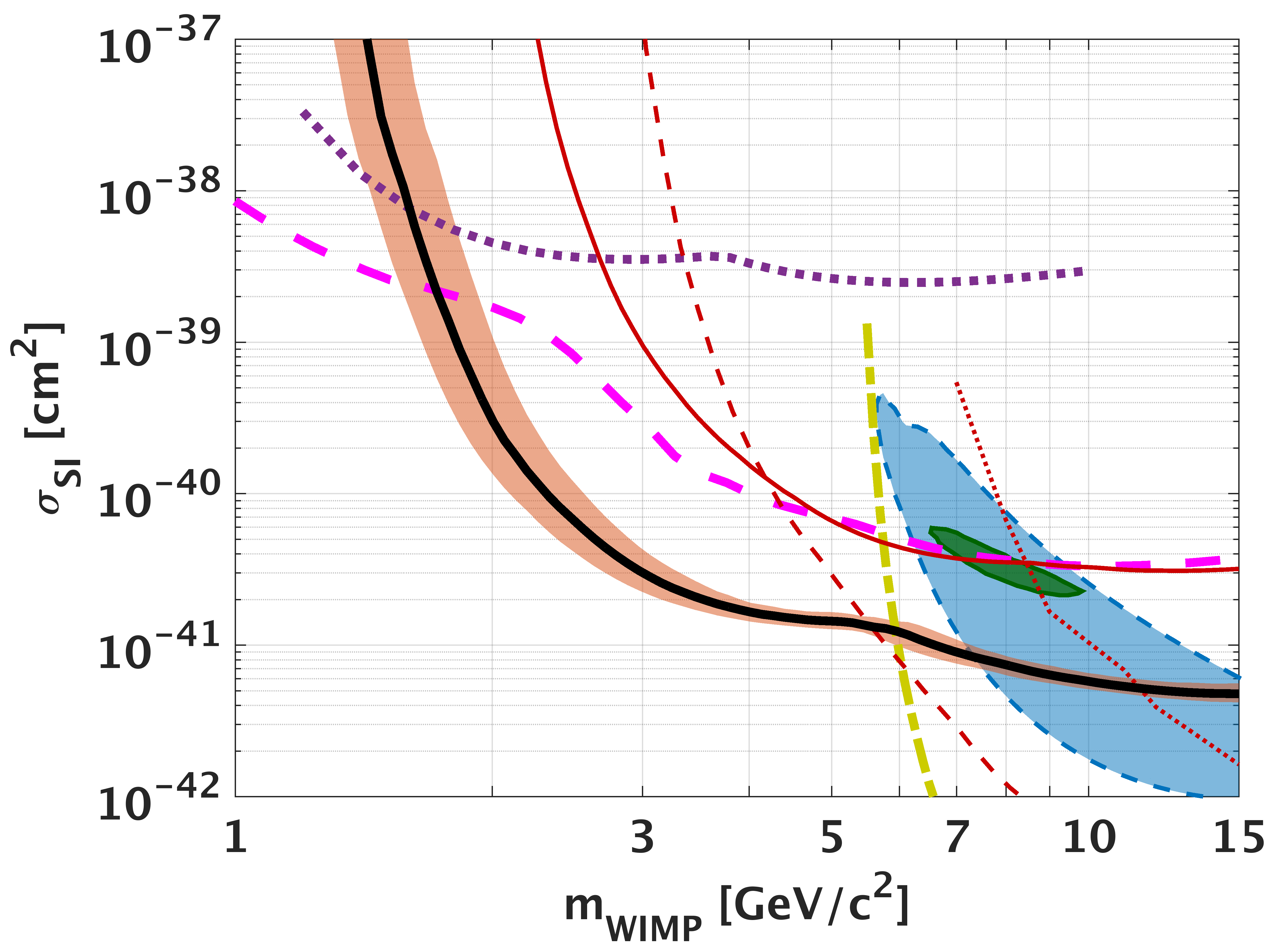}
		\caption{(color online) Median (90\,\% C.L.) and 95\,\% interval of the WIMP limit from this analysis (black thick solid surrounded by salmon-shaded band) compared to other selected results.  Other 90\,\% upper limits shown are from the first CDMSlite run (red thin solid curve)~\cite{Agnese2014}, SuperCDMS (red thin dashed curve)~\cite{Agnese2014a}, EDELWEISS-II (red thin dotted curve)~\cite{Armengaud2012}, LUX (dark-yellow thick dashed-dot curve)~\cite{Akerib2014}, CRESST (magenta thick dashed curve)~\cite{Angloher2015}, and DAMIC (purple thick dotted curve)~\cite{Barreto2012}.  Closed regions are CDMS II Si 90\,\% C.L. (blue dashed shaded region)~\cite{Agnese2013}, and CoGeNT 90\,\% C.L. (dark-green shaded region)~\cite{Aalseth2013}.}
		\label{fig:limit}
\end{figure}

The data below 2~\keVee and above the respective 50\,\% trigger-efficiency values were used to set the 90\% confidence upper limit on the spin-independent WIMP-nucleon cross section.  This was done using the optimum interval method with no background subtraction~\cite{Yellin2002,*Yellin2007}, the Helm form factor, and the following standard dark matter halo assumptions: a local dark matter density of 0.3~$\text{GeV}\,\text{cm}^{-3}$, a most probable Galactic WIMP velocity of $220~\text{km}\,\text{s}^{-1}$, a mean orbital velocity of Earth with respect to the Galactic Center of $232~\text{km}\,\text{s}^{-1}$, and a Galactic escape velocity of $544~\text{km}\,\text{s}^{-1}$~\cite{Lewin1996}.  The results are shown in \sfig\ref{fig:limit}.

Statistical and systematic uncertainties were propagated into the limit by calculating the final efficiency as a function of energy numerous times, each time picking at random from the distributions of each input parameter.  Statistical uncertainties exist in the trigger efficiency, pulse-shape and radial-cut simulations, and energy-independent cuts, while systematic uncertainties exist in the pulse-shape and radial-cut simulations, and the Lindhard model $k$-value.
Limits were computed using 1000 sample curves with the median and 95\,\% interval given in \sfig\ref{fig:limit}.  The uncertainty is dominated by the Lindhard model, particularly below masses of 3~\gev, and the radial-cut efficiency.

This result excludes new parameter space for WIMP masses between 1.6 and 5.5~\gev.  The improvement in sensitivity over the first CDMSlite run is due to the increase in exposure, the reduction in threshold, and the decrease in background resulting from the radial fiducial-volume cut.  A kink in the limit is seen at ${\sim}6$~\gev.  Simulations indicate that this feature is a consequence of the $M$-shell line at 160~\eVeeNS.  Finally, the effect of having a tighter radial threshold in period 2 was considered.  Placing the same looser threshold in both periods would result in a ${\sim}$9\,\% weakened sensitivity below masses of 6~\gev, which is well within the presented uncertainty band.

In conclusion, the second CDMSlite run was successful in operating an iZIP detector at a bias potential of $-70$~V for a 70.10~\kgd analysis exposure, with ionization thresholds of 75 and 56~\eVee attained for the first and second period, respectively.  The development of a fiducial-volume cut reduced the overall background rate significantly.  The results presented here can be significantly improved in future CDMSlite runs by lowering the threshold and background rate.  The former can be achieved with better phonon resolution and higher bias potentials, and the latter with material selection and quality control.  All of these improvements are planned for the SuperCDMS SNOLAB experiment.

The SuperCDMS collaboration gratefully acknowledges technical assistance from the staff of the Soudan Underground Laboratory and the Minnesota Department of Natural Resources. The iZIP detectors were fabricated in the Stanford Nanofabrication Facility, which is a member of the National Nanofabrication Infrastructure Network, sponsored and supported by the NSF. Part of the research described in this Letter was conducted under the Ultra Sensitive Nuclear Measurements Initiative at Pacific Northwest National Laboratory, which is operated by Battelle for the U.S. Department of Energy. Funding and support were received from the National Science Foundation, the Department of Energy, Fermilab URA Visiting Scholar Award 13-S-04, NSERC Canada, and MultiDark (Spanish MINECO). Fermilab is operated by the Fermi Research Alliance, LLC under Contract No. De-AC02-07CH11359. SLAC is operated under Contract No. DEAC02-76SF00515 with the United States Department of Energy.

\bibliography{CDMSliteR2}

\begin{thebibliography}{56}%
\makeatletter
\providecommand \@ifxundefined [1]{%
 \@ifx{#1\undefined}
}%
\providecommand \@ifnum [1]{%
 \ifnum #1\expandafter \@firstoftwo
 \else \expandafter \@secondoftwo
 \fi
}%
\providecommand \@ifx [1]{%
 \ifx #1\expandafter \@firstoftwo
 \else \expandafter \@secondoftwo
 \fi
}%
\providecommand \natexlab [1]{#1}%
\providecommand \enquote  [1]{``#1''}%
\providecommand \bibnamefont  [1]{#1}%
\providecommand \bibfnamefont [1]{#1}%
\providecommand \citenamefont [1]{#1}%
\providecommand \href@noop [0]{\@secondoftwo}%
\providecommand \href [0]{\begingroup \@sanitize@url \@href}%
\providecommand \@href[1]{\@@startlink{#1}\@@href}%
\providecommand \@@href[1]{\endgroup#1\@@endlink}%
\providecommand \@sanitize@url [0]{\catcode `\\12\catcode `\$12\catcode
  `\&12\catcode `\#12\catcode `\^12\catcode `\_12\catcode `\%12\relax}%
\providecommand \@@startlink[1]{}%
\providecommand \@@endlink[0]{}%
\providecommand \url  [0]{\begingroup\@sanitize@url \@url }%
\providecommand \@url [1]{\endgroup\@href {#1}{\urlprefix }}%
\providecommand \urlprefix  [0]{URL }%
\providecommand \Eprint [0]{\href }%
\providecommand \doibase [0]{http://dx.doi.org/}%
\providecommand \selectlanguage [0]{\@gobble}%
\providecommand \bibinfo  [0]{\@secondoftwo}%
\providecommand \bibfield  [0]{\@secondoftwo}%
\providecommand \translation [1]{[#1]}%
\providecommand \BibitemOpen [0]{}%
\providecommand \bibitemStop [0]{}%
\providecommand \bibitemNoStop [0]{.\EOS\space}%
\providecommand \EOS [0]{\spacefactor3000\relax}%
\providecommand \BibitemShut  [1]{\csname bibitem#1\endcsname}%
\let\auto@bib@innerbib\@empty
\bibitem [{\citenamefont {Olive}\ \emph {et~al.}(2014)\citenamefont {Olive}
  \emph {et~al.}}]{Olive2014}%
  \BibitemOpen
  \bibfield  {author} {\bibinfo {author} {\bibfnamefont {K.}~\bibnamefont
  {Olive}}\  \emph {et~al.} (\bibinfo {collaboration} {Particle Data Group}),\
  }\href {\doibase 10.1088/1674-1137/38/9/090001} {\bibfield  {journal}
  {\bibinfo  {journal} {Chin. Phys. C}\ }\textbf {\bibinfo {volume} {38}},\
  \bibinfo {pages} {090001} (\bibinfo {year} {2014})}\BibitemShut {NoStop}%
\bibitem [{\citenamefont {{Ade, P. A. R.}}\ \emph {et~al.}(2014)\citenamefont
  {{Ade, P. A. R.}} \emph {et~al.}}]{Ade2014}%
  \BibitemOpen
  \bibfield  {author} {\bibinfo {author} {\bibnamefont {{Ade, P. A. R.}}}\
  \emph {et~al.} (\bibinfo {collaboration} {Planck Collaboration}),\ }\href
  {\doibase 10.1051/0004-6361/201321591} {\bibfield  {journal} {\bibinfo
  {journal} {Astron. Astrophys.}\ }\textbf {\bibinfo {volume} {571}},\ \bibinfo
  {pages} {A16} (\bibinfo {year} {2014})}\BibitemShut {NoStop}%
\bibitem [{\citenamefont {Jungman}\ \emph {et~al.}(1996)\citenamefont
  {Jungman}, \citenamefont {Kamionkowski},\ and\ \citenamefont
  {Griest}}]{Jungman1996}%
  \BibitemOpen
  \bibfield  {author} {\bibinfo {author} {\bibfnamefont {G.}~\bibnamefont
  {Jungman}}, \bibinfo {author} {\bibfnamefont {M.}~\bibnamefont
  {Kamionkowski}}, and\ \bibinfo {author} {\bibfnamefont {K.}~\bibnamefont
  {Griest}},\ }\href {\doibase 10.1016/0370-1573(95)00058-5} {\bibfield
  {journal} {\bibinfo  {journal} {Phys. Rep.}\ }\textbf {\bibinfo {volume}
  {267}},\ \bibinfo {pages} {195} (\bibinfo {year} {1996})}\BibitemShut
  {NoStop}%
\bibitem [{\citenamefont {Abercrombie}\ \emph {et~al.}(2015)\citenamefont
  {Abercrombie} \emph {et~al.}}]{Abercrombie2015}%
  \BibitemOpen
  \bibfield  {author} {\bibinfo {author} {\bibfnamefont {D.}~\bibnamefont
  {Abercrombie}}\  \emph {et~al.},\ }\href@noop {} {}\Eprint
  {http://arxiv.org/abs/1507.00966} {arXiv:1507.00966 [hep-ex]} \BibitemShut
  {NoStop}%
\bibitem [{\citenamefont {Akerib}\ \emph {et~al.}(2014)\citenamefont {Akerib}
  \emph {et~al.}}]{Akerib2014}%
  \BibitemOpen
  \bibfield  {author} {\bibinfo {author} {\bibfnamefont {D.}~\bibnamefont
  {Akerib}}\  \emph {et~al.} (\bibinfo {collaboration} {LUX Collaboration}),\
  }\href {\doibase 10.1103/PhysRevLett.112.091303} {\bibfield  {journal}
  {\bibinfo  {journal} {Phys. Rev. Lett.}\ }\textbf {\bibinfo {volume} {112}},\
  \bibinfo {pages} {091303} (\bibinfo {year} {2014})}\BibitemShut {NoStop}%
\bibitem [{\citenamefont {Hooper}\ \emph {et~al.}(2012)\citenamefont {Hooper},
  \citenamefont {Weiner},\ and\ \citenamefont {Xue}}]{PhysRevD.86.056009}%
  \BibitemOpen
  \bibfield  {author} {\bibinfo {author} {\bibfnamefont {D.}~\bibnamefont
  {Hooper}}, \bibinfo {author} {\bibfnamefont {N.}~\bibnamefont {Weiner}}, and\
  \bibinfo {author} {\bibfnamefont {W.}~\bibnamefont {Xue}},\ }\href {\doibase
  10.1103/PhysRevD.86.056009} {\bibfield  {journal} {\bibinfo  {journal} {Phys.
  Rev. D}\ }\textbf {\bibinfo {volume} {86}},\ \bibinfo {pages} {056009}
  (\bibinfo {year} {2012})}\BibitemShut {NoStop}%
\bibitem [{\citenamefont {Kusenko}(2009)}]{Kusenko20091}%
  \BibitemOpen
  \bibfield  {author} {\bibinfo {author} {\bibfnamefont {A.}~\bibnamefont
  {Kusenko}},\ }\href {\doibase
  http://dx.doi.org/10.1016/j.physrep.2009.07.004} {\bibfield  {journal}
  {\bibinfo  {journal} {Physics Reports}\ }\textbf {\bibinfo {volume} {481}},\
  \bibinfo {pages} {1 } (\bibinfo {year} {2009})}\BibitemShut {NoStop}%
\bibitem [{\citenamefont {Pospelov}\ \emph {et~al.}(2008)\citenamefont
  {Pospelov}, \citenamefont {Ritz},\ and\ \citenamefont
  {Voloshin}}]{PhysRevD.78.115012}%
  \BibitemOpen
  \bibfield  {author} {\bibinfo {author} {\bibfnamefont {M.}~\bibnamefont
  {Pospelov}}, \bibinfo {author} {\bibfnamefont {A.}~\bibnamefont {Ritz}}, and\
  \bibinfo {author} {\bibfnamefont {M.}~\bibnamefont {Voloshin}},\ }\href
  {\doibase 10.1103/PhysRevD.78.115012} {\bibfield  {journal} {\bibinfo
  {journal} {Phys. Rev. D}\ }\textbf {\bibinfo {volume} {78}},\ \bibinfo
  {pages} {115012} (\bibinfo {year} {2008})}\BibitemShut {NoStop}%
\bibitem [{\citenamefont {Fayet}(2007)}]{PhysRevD.75.115017}%
  \BibitemOpen
  \bibfield  {author} {\bibinfo {author} {\bibfnamefont {P.}~\bibnamefont
  {Fayet}},\ }\href {\doibase 10.1103/PhysRevD.75.115017} {\bibfield  {journal}
  {\bibinfo  {journal} {Phys. Rev. D}\ }\textbf {\bibinfo {volume} {75}},\
  \bibinfo {pages} {115017} (\bibinfo {year} {2007})}\BibitemShut {NoStop}%
\bibitem [{\citenamefont {Feng}\ and\ \citenamefont
  {Kumar}(2008)}]{PhysRevLett.101.231301}%
  \BibitemOpen
  \bibfield  {author} {\bibinfo {author} {\bibfnamefont {J.~L.}\ \bibnamefont
  {Feng}}\ and\ \bibinfo {author} {\bibfnamefont {J.}~\bibnamefont {Kumar}},\
  }\href {\doibase 10.1103/PhysRevLett.101.231301} {\bibfield  {journal}
  {\bibinfo  {journal} {Phys. Rev. Lett.}\ }\textbf {\bibinfo {volume} {101}},\
  \bibinfo {pages} {231301} (\bibinfo {year} {2008})}\BibitemShut {NoStop}%
\bibitem [{\citenamefont {Boehm}\ and\ \citenamefont
  {Fayet}(2004)}]{Boehm2004219}%
  \BibitemOpen
  \bibfield  {author} {\bibinfo {author} {\bibfnamefont {C.}~\bibnamefont
  {Boehm}}\ and\ \bibinfo {author} {\bibfnamefont {P.}~\bibnamefont {Fayet}},\
  }\href {\doibase http://dx.doi.org/10.1016/j.nuclphysb.2004.01.015}
  {\bibfield  {journal} {\bibinfo  {journal} {Nucl. Phys.}\ }\textbf {\bibinfo
  {volume} {683B}},\ \bibinfo {pages} {219 } (\bibinfo {year}
  {2004})}\BibitemShut {NoStop}%
\bibitem [{\citenamefont {Kaplan}(1992)}]{Kaplan1992}%
  \BibitemOpen
  \bibfield  {author} {\bibinfo {author} {\bibfnamefont {D.~B.}\ \bibnamefont
  {Kaplan}},\ }\href {\doibase 10.1103/PhysRevLett.68.741} {\bibfield
  {journal} {\bibinfo  {journal} {Phys. Rev. Lett.}\ }\textbf {\bibinfo
  {volume} {68}},\ \bibinfo {pages} {741} (\bibinfo {year} {1992})}\BibitemShut
  {NoStop}%
\bibitem [{\citenamefont {Kaplan}\ \emph {et~al.}(2009)\citenamefont {Kaplan},
  \citenamefont {Luty},\ and\ \citenamefont {Zurek}}]{Kaplan2009}%
  \BibitemOpen
  \bibfield  {author} {\bibinfo {author} {\bibfnamefont {D.~E.}\ \bibnamefont
  {Kaplan}}, \bibinfo {author} {\bibfnamefont {M.~A.}\ \bibnamefont {Luty}},
  and\ \bibinfo {author} {\bibfnamefont {K.~M.}\ \bibnamefont {Zurek}},\ }\href
  {\doibase 10.1103/PhysRevD.79.115016} {\bibfield  {journal} {\bibinfo
  {journal} {Phys. Rev. D}\ }\textbf {\bibinfo {volume} {79}},\ \bibinfo
  {pages} {115016} (\bibinfo {year} {2009})}\BibitemShut {NoStop}%
\bibitem [{\citenamefont {Falkowski}\ \emph {et~al.}(2011)\citenamefont
  {Falkowski}, \citenamefont {Ruderman},\ and\ \citenamefont
  {Volansky}}]{Falkowski2011}%
  \BibitemOpen
  \bibfield  {author} {\bibinfo {author} {\bibfnamefont {A.}~\bibnamefont
  {Falkowski}}, \bibinfo {author} {\bibfnamefont {J.}~\bibnamefont {Ruderman}},
  and\ \bibinfo {author} {\bibfnamefont {T.}~\bibnamefont {Volansky}},\ }\href
  {\doibase 10.1007/JHEP05(2011)106} {\bibfield  {journal} {\bibinfo  {journal}
  {J. High Energy Phys.}\ }\bibinfo {volume} {05} (\bibinfo {year} {2011})\
  \bibinfo {pages} {106}}\BibitemShut {NoStop}%
\bibitem [{\citenamefont {Volkas}\ and\ \citenamefont
  {Petraki}(2013)}]{Volkas2013}%
  \BibitemOpen
  \bibfield  {author} {\bibinfo {author} {\bibfnamefont {R.~R.}\ \bibnamefont
  {Volkas}}\ and\ \bibinfo {author} {\bibfnamefont {K.}~\bibnamefont
  {Petraki}},\ }\href {\doibase 10.1142/S0217751X13300287} {\bibfield
  {journal} {\bibinfo  {journal} {Int. J. Mod. Phys. A}\ }\textbf {\bibinfo
  {volume} {28}},\ \bibinfo {pages} {1330028} (\bibinfo {year}
  {2013})}\BibitemShut {NoStop}%
\bibitem [{\citenamefont {{Zurek}}(2014)}]{Zurek2014}%
  \BibitemOpen
  \bibfield  {author} {\bibinfo {author} {\bibfnamefont {K.~M.}\ \bibnamefont
  {{Zurek}}},\ }\href {\doibase 10.1016/j.physrep.2013.12.001} {\bibfield
  {journal} {\bibinfo  {journal} {Phys. Rep.}\ }\textbf {\bibinfo {volume}
  {537}},\ \bibinfo {pages} {91} (\bibinfo {year} {2014})}\BibitemShut
  {NoStop}%
\bibitem [{\citenamefont {Agnese}\ \emph
  {et~al.}(2013{\natexlab{a}})\citenamefont {Agnese} \emph
  {et~al.}}]{Agnese2013}%
  \BibitemOpen
  \bibfield  {author} {\bibinfo {author} {\bibfnamefont {R.}~\bibnamefont
  {Agnese}}\  \emph {et~al.} (\bibinfo {collaboration} {SuperCDMS
  Collaboration}),\ }\href {\doibase 10.1103/PhysRevLett.111.251301} {\bibfield
   {journal} {\bibinfo  {journal} {Phys. Rev. Lett.}\ }\textbf {\bibinfo
  {volume} {111}},\ \bibinfo {pages} {251301} (\bibinfo {year}
  {2013}{\natexlab{a}})}\BibitemShut {NoStop}%
\bibitem [{\citenamefont {Angloher}\ \emph {et~al.}(2012)\citenamefont
  {Angloher} \emph {et~al.}}]{Angloher2012}%
  \BibitemOpen
  \bibfield  {author} {\bibinfo {author} {\bibfnamefont {G.}~\bibnamefont
  {Angloher}}\  \emph {et~al.} (\bibinfo {collaboration} {CRESST
  Collaboration}),\ }\href {\doibase 10.1140/epjc/s10052-012-1971-8} {\bibfield
   {journal} {\bibinfo  {journal} {Eur. Phys. J. C}\ }\textbf {\bibinfo
  {volume} {72}},\ \bibinfo {pages} {1971} (\bibinfo {year}
  {2012})}\BibitemShut {NoStop}%
\bibitem [{\citenamefont {Aalseth}\ \emph {et~al.}(2013)\citenamefont
  {Aalseth}, \citenamefont {Barbeau}, \citenamefont {Colaresi}, \citenamefont
  {Collar}, \citenamefont {DiazLeon}, \citenamefont {Fast}, \citenamefont
  {Fields}, \citenamefont {Hossbach}, \citenamefont {Knecht}, \citenamefont
  {Kos}, \citenamefont {Marino}, \citenamefont {Miley}, \citenamefont {Miller},
  \citenamefont {Orrell}, ,\ and\ \citenamefont {Yocum}}]{Aalseth2013}%
  \BibitemOpen
  \bibfield  {author} {\bibinfo {author} {\bibfnamefont {C.~E.}\ \bibnamefont
  {Aalseth}}, \bibinfo {author} {\bibfnamefont {P.~S.}\ \bibnamefont
  {Barbeau}}, \bibinfo {author} {\bibfnamefont {J.}~\bibnamefont {Colaresi}},
  \bibinfo {author} {\bibfnamefont {J.~I.}\ \bibnamefont {Collar}}, \bibinfo
  {author} {\bibfnamefont {J.}~\bibnamefont {DiazLeon}}, \bibinfo {author}
  {\bibfnamefont {J.~E.}\ \bibnamefont {Fast}}, \bibinfo {author}
  {\bibfnamefont {N.~E.}\ \bibnamefont {Fields}}, \bibinfo {author}
  {\bibfnamefont {T.}~\bibnamefont {Hossbach}}, \bibinfo {author}
  {\bibfnamefont {A.}~\bibnamefont {Knecht}}, \bibinfo {author} {\bibfnamefont
  {M.~S.}\ \bibnamefont {Kos}}, \bibinfo {author} {\bibfnamefont {M.~G.}\
  \bibnamefont {Marino}}, \bibinfo {author} {\bibfnamefont {H.~S.}\
  \bibnamefont {Miley}}, \bibinfo {author} {\bibfnamefont {M.~L.}\ \bibnamefont
  {Miller}}, \bibinfo {author} {\bibfnamefont {J.~L.}\ \bibnamefont {Orrell}},
  , and\ \bibinfo {author} {\bibfnamefont {K.~M.}\ \bibnamefont {Yocum}}
  (\bibinfo {collaboration} {CoGeNT Collaboration}),\ }\href {\doibase
  10.1103/PhysRevD.88.012002} {\bibfield  {journal} {\bibinfo  {journal} {Phys.
  Rev. D}\ }\textbf {\bibinfo {volume} {88}},\ \bibinfo {pages} {012002}
  (\bibinfo {year} {2013})}\BibitemShut {NoStop}%
\bibitem [{\citenamefont {Savage}\ \emph {et~al.}(2009)\citenamefont {Savage},
  \citenamefont {Gelmini}, \citenamefont {Gondolo},\ and\ \citenamefont
  {Freese}}]{Savage2009}%
  \BibitemOpen
  \bibfield  {author} {\bibinfo {author} {\bibfnamefont {C.}~\bibnamefont
  {Savage}}, \bibinfo {author} {\bibfnamefont {G.}~\bibnamefont {Gelmini}},
  \bibinfo {author} {\bibfnamefont {P.}~\bibnamefont {Gondolo}}, and\ \bibinfo
  {author} {\bibfnamefont {K.}~\bibnamefont {Freese}},\ }\href {\doibase
  10.1088/1475-7516/2009/04/010} {\bibfield  {journal} {\bibinfo  {journal} {J.
  Cosmol. Astropart. Phys.}\ }\bibinfo {volume} {04} (\bibinfo {year} {2009})\
  \bibinfo {pages} {010}}\BibitemShut {NoStop}%
\bibitem [{\citenamefont {Bernabei}\ \emph {et~al.}(2010)\citenamefont
  {Bernabei} \emph {et~al.}}]{Bernabei2010}%
  \BibitemOpen
  \bibfield  {author} {\bibinfo {author} {\bibfnamefont {R.}~\bibnamefont
  {Bernabei}}\  \emph {et~al.} (\bibinfo {collaboration} {DAMA/LIBRA
  Collaboration}),\ }\href {\doibase 10.1140/epjc/s10052-010-1303-9} {\bibfield
   {journal} {\bibinfo  {journal} {Eur. Phys. J. C}\ }\textbf {\bibinfo
  {volume} {67}},\ \bibinfo {pages} {39} (\bibinfo {year} {2010})}\BibitemShut
  {NoStop}%
\bibitem [{\citenamefont {{Hooper}}\ and\ \citenamefont
  {{Goodenough}}(2011)}]{Hooper2011}%
  \BibitemOpen
  \bibfield  {author} {\bibinfo {author} {\bibfnamefont {D.}~\bibnamefont
  {{Hooper}}}\ and\ \bibinfo {author} {\bibfnamefont {L.}~\bibnamefont
  {{Goodenough}}},\ }\href {\doibase 10.1016/j.physletb.2011.02.029} {\bibfield
   {journal} {\bibinfo  {journal} {Phys. Lett.}\ }\textbf {\bibinfo {volume}
  {697}},\ \bibinfo {pages} {412} (\bibinfo {year} {2011})}\BibitemShut
  {NoStop}%
\bibitem [{\citenamefont {Agnese}\ \emph
  {et~al.}(2014{\natexlab{a}})\citenamefont {Agnese} \emph
  {et~al.}}]{Agnese2014}%
  \BibitemOpen
  \bibfield  {author} {\bibinfo {author} {\bibfnamefont {R.}~\bibnamefont
  {Agnese}}\  \emph {et~al.} (\bibinfo {collaboration} {SuperCDMS
  Collaboration}),\ }\href {\doibase 10.1103/PhysRevLett.112.041302} {\bibfield
   {journal} {\bibinfo  {journal} {Phys. Rev. Lett.}\ }\textbf {\bibinfo
  {volume} {112}},\ \bibinfo {pages} {041302} (\bibinfo {year}
  {2014}{\natexlab{a}})}\BibitemShut {NoStop}%
\bibitem [{\citenamefont {Agnese}\ \emph
  {et~al.}(2014{\natexlab{b}})\citenamefont {Agnese} \emph
  {et~al.}}]{Agnese2014a}%
  \BibitemOpen
  \bibfield  {author} {\bibinfo {author} {\bibfnamefont {R.}~\bibnamefont
  {Agnese}}\  \emph {et~al.} (\bibinfo {collaboration} {SuperCDMS
  Collaboration}),\ }\href {\doibase 10.1103/PhysRevLett.112.241302} {\bibfield
   {journal} {\bibinfo  {journal} {Phys. Rev. Lett.}\ }\textbf {\bibinfo
  {volume} {112}},\ \bibinfo {pages} {241302} (\bibinfo {year}
  {2014}{\natexlab{b}})}\BibitemShut {NoStop}%
\bibitem [{\citenamefont {Armengaud}\ \emph {et~al.}(2012)\citenamefont
  {Armengaud} \emph {et~al.}}]{Armengaud2012}%
  \BibitemOpen
  \bibfield  {author} {\bibinfo {author} {\bibfnamefont {E.}~\bibnamefont
  {Armengaud}}\  \emph {et~al.} (\bibinfo {collaboration} {EDELWEISS-II
  Collaboration}),\ }\href {\doibase 10.1103/PhysRevD.86.051701} {\bibfield
  {journal} {\bibinfo  {journal} {Phys. Rev. D}\ }\textbf {\bibinfo {volume}
  {86}},\ \bibinfo {pages} {051701} (\bibinfo {year} {2012})}\BibitemShut
  {NoStop}%
\bibitem [{\citenamefont {Angloher}\ \emph {et~al.}(2014)\citenamefont
  {Angloher} \emph {et~al.}}]{Angloher2014}%
  \BibitemOpen
  \bibfield  {author} {\bibinfo {author} {\bibfnamefont {G.}~\bibnamefont
  {Angloher}}\  \emph {et~al.} (\bibinfo {collaboration} {CRESST
  Collaboration}),\ }\href {\doibase 10.1140/epjc/s10052-014-3184-9} {\bibfield
   {journal} {\bibinfo  {journal} {Eur. Phys. J. C}\ }\textbf {\bibinfo
  {volume} {74}},\ \bibinfo {pages} {3184} (\bibinfo {year}
  {2014})}\BibitemShut {NoStop}%
\bibitem [{\citenamefont {Angloher}\ \emph {et~al.}(2016)\citenamefont
  {Angloher} \emph {et~al.}}]{Angloher2015}%
  \BibitemOpen
  \bibfield  {author} {\bibinfo {author} {\bibfnamefont {G.}~\bibnamefont
  {Angloher}}\  \emph {et~al.} (\bibinfo {collaboration} {CRESST
  Collaboration}),\ }\href@noop {} {}\bibinfo {note} {Accepted to Eur. Phys. J.
  C},\ \Eprint {http://arxiv.org/abs/1509.01515v1} {arXiv:1509.01515v1
  [astro-ph.CO]} \BibitemShut {NoStop}%
\bibitem [{\citenamefont {Barreto}\ \emph {et~al.}(2012)\citenamefont {Barreto}
  \emph {et~al.}}]{Barreto2012}%
  \BibitemOpen
  \bibfield  {author} {\bibinfo {author} {\bibfnamefont {J.}~\bibnamefont
  {Barreto}}\  \emph {et~al.} (\bibinfo {collaboration} {DAMIC
  Collaboration}),\ }\href {\doibase 10.1016/j.physletb.2012.04.006} {\bibfield
   {journal} {\bibinfo  {journal} {Phys. Lett. B}\ }\textbf {\bibinfo {volume}
  {711}},\ \bibinfo {pages} {264} (\bibinfo {year} {2012})}\BibitemShut
  {NoStop}%
\bibitem [{\citenamefont {Amole}\ \emph {et~al.}(2015)\citenamefont {Amole}
  \emph {et~al.}}]{Amole2015}%
  \BibitemOpen
  \bibfield  {author} {\bibinfo {author} {\bibfnamefont {C.}~\bibnamefont
  {Amole}}\  \emph {et~al.} (\bibinfo {collaboration} {PICO Collaboration}),\
  }\href {\doibase 10.1103/PhysRevLett.114.231302} {\bibfield  {journal}
  {\bibinfo  {journal} {Phys. Rev. Lett.}\ }\textbf {\bibinfo {volume} {114}},\
  \bibinfo {pages} {231302} (\bibinfo {year} {2015})}\BibitemShut {NoStop}%
\bibitem [{\citenamefont {Yue}\ \emph {et~al.}(2014)\citenamefont {Yue} \emph
  {et~al.}}]{Yue2014}%
  \BibitemOpen
  \bibfield  {author} {\bibinfo {author} {\bibfnamefont {Q.}~\bibnamefont
  {Yue}}\  \emph {et~al.} (\bibinfo {collaboration} {CDEX Collaboration}),\
  }\href {\doibase 10.1103/PhysRevD.90.091701} {\bibfield  {journal} {\bibinfo
  {journal} {Phys. Rev. D}\ }\textbf {\bibinfo {volume} {90}},\ \bibinfo
  {pages} {091701(R)} (\bibinfo {year} {2014})}\BibitemShut {NoStop}%
\bibitem [{\citenamefont {Li}\ \emph {et~al.}(2013)\citenamefont {Li} \emph
  {et~al.}}]{Li2013}%
  \BibitemOpen
  \bibfield  {author} {\bibinfo {author} {\bibfnamefont {H.~B.}\ \bibnamefont
  {Li}}\  \emph {et~al.} (\bibinfo {collaboration} {TEXONO Collaboration}),\
  }\href {\doibase 10.1103/PhysRevLett.110.261301} {\bibfield  {journal}
  {\bibinfo  {journal} {Phys. Rev. Lett.}\ }\textbf {\bibinfo {volume} {110}},\
  \bibinfo {pages} {261301} (\bibinfo {year} {2013})}\BibitemShut {NoStop}%
\bibitem [{\citenamefont {Angle}\ \emph {et~al.}(2011)\citenamefont {Angle}
  \emph {et~al.}}]{Angle2011}%
  \BibitemOpen
  \bibfield  {author} {\bibinfo {author} {\bibfnamefont {J.}~\bibnamefont
  {Angle}}\  \emph {et~al.} (\bibinfo {collaboration} {XENON10
  Collaboration}),\ }\href {\doibase 10.1103/PhysRevLett.107.051301} {\bibfield
   {journal} {\bibinfo  {journal} {Phys. Rev. Lett.}\ }\textbf {\bibinfo
  {volume} {107}},\ \bibinfo {pages} {051301} (\bibinfo {year}
  {2011})}\BibitemShut {NoStop}%
\bibitem [{\citenamefont {Angle}\ \emph {et~al.}(2013)\citenamefont {Angle}
  \emph {et~al.}}]{Angle2013}%
  \BibitemOpen
  \bibfield  {author} {\bibinfo {author} {\bibfnamefont {J.}~\bibnamefont
  {Angle}}\  \emph {et~al.} (\bibinfo {collaboration} {XENON10
  Collaboration}),\ }\href {\doibase 10.1103/PhysRevLett.110.249901} {\bibfield
   {journal} {\bibinfo  {journal} {Phys. Rev. Lett.}\ }\textbf {\bibinfo
  {volume} {110}},\ \bibinfo {pages} {249901(E)} (\bibinfo {year}
  {2013})}\BibitemShut {NoStop}%
\bibitem [{\citenamefont {Goodman}\ and\ \citenamefont
  {Witten}(1985)}]{Goodman1985}%
  \BibitemOpen
  \bibfield  {author} {\bibinfo {author} {\bibfnamefont {M.~W.}\ \bibnamefont
  {Goodman}}\ and\ \bibinfo {author} {\bibfnamefont {E.}~\bibnamefont
  {Witten}},\ }\href {\doibase 10.1103/PhysRevD.31.3059} {\bibfield  {journal}
  {\bibinfo  {journal} {Phys. Rev. D}\ }\textbf {\bibinfo {volume} {31}},\
  \bibinfo {pages} {3059} (\bibinfo {year} {1985})}\BibitemShut {NoStop}%
\bibitem [{\citenamefont {Agnese}\ \emph
  {et~al.}(2013{\natexlab{b}})\citenamefont {Agnese} \emph
  {et~al.}}]{Agnese2013a}%
  \BibitemOpen
  \bibfield  {author} {\bibinfo {author} {\bibfnamefont {R.}~\bibnamefont
  {Agnese}}\  \emph {et~al.} (\bibinfo {collaboration} {SuperCDMS
  Collaboration}),\ }\href {\doibase 10.1063/1.4826093} {\bibfield  {journal}
  {\bibinfo  {journal} {Appl. Phys. Lett.}\ }\textbf {\bibinfo {volume}
  {103}},\ \bibinfo {pages} {164105} (\bibinfo {year}
  {2013}{\natexlab{b}})}\BibitemShut {NoStop}%
\bibitem [{\citenamefont {Akerib}\ \emph {et~al.}(2005)\citenamefont {Akerib}
  \emph {et~al.}}]{Akerib2005}%
  \BibitemOpen
  \bibfield  {author} {\bibinfo {author} {\bibfnamefont {D.~S.}\ \bibnamefont
  {Akerib}}\  \emph {et~al.} (\bibinfo {collaboration} {CDMS Collaboration}),\
  }\href {\doibase 10.1103/PhysRevD.72.052009} {\bibfield  {journal} {\bibinfo
  {journal} {Phys. Rev. D}\ }\textbf {\bibinfo {volume} {72}},\ \bibinfo
  {pages} {052009} (\bibinfo {year} {2005})}\BibitemShut {NoStop}%
\bibitem [{\citenamefont {Luke}(1988)}]{Luke1988}%
  \BibitemOpen
  \bibfield  {author} {\bibinfo {author} {\bibfnamefont {P.~N.}\ \bibnamefont
  {Luke}},\ }\href {\doibase 10.1063/1.341976} {\bibfield  {journal} {\bibinfo
  {journal} {J. Appl. Phys.}\ }\textbf {\bibinfo {volume} {64}},\ \bibinfo
  {pages} {6858} (\bibinfo {year} {1988})}\BibitemShut {NoStop}%
\bibitem [{\citenamefont {Neganov}\ and\ \citenamefont
  {Trofimov}(1985)}]{Neganov1985}%
  \BibitemOpen
  \bibfield  {author} {\bibinfo {author} {\bibfnamefont {B.~S.}\ \bibnamefont
  {Neganov}}\ and\ \bibinfo {author} {\bibfnamefont {V.~N.}\ \bibnamefont
  {Trofimov}},\ }\href
  {http://patents.su/3-1037771-sposob-kalorimetricheskogo-izmereniya-ioniziruyushhikh-izluchenijj.html}
  {\bibfield  {journal} {\bibinfo  {journal} {Otkrytia i izobreteniya}\ ,\
  \bibinfo {pages} {215}} (\bibinfo {year} {1985})},\ \bibinfo {note} {{USSR
  Patent No 1037771}}\BibitemShut {NoStop}%
\bibitem [{\citenamefont {Wang}(2010)}]{Wang2010}%
  \BibitemOpen
  \bibfield  {author} {\bibinfo {author} {\bibfnamefont {G.}~\bibnamefont
  {Wang}},\ }\href {\doibase 10.1063/1.3354095} {\bibfield  {journal} {\bibinfo
   {journal} {J. Appl. Phys.}\ }\textbf {\bibinfo {volume} {107}},\ \bibinfo
  {pages} {094504} (\bibinfo {year} {2010})}\BibitemShut {NoStop}%
\bibitem [{\citenamefont {Akerib}\ \emph {et~al.}(2004)\citenamefont {Akerib},
  \citenamefont {Dragowsky}, \citenamefont {Driscoll}, \citenamefont {Kamat},
  \citenamefont {Perera}, \citenamefont {Schnee}, \citenamefont {Wang},
  \citenamefont {Gaitskell}, \citenamefont {Bogdanova},\ and\ \citenamefont
  {Trofimov}}]{Akerib2004}%
  \BibitemOpen
  \bibfield  {author} {\bibinfo {author} {\bibfnamefont {D.}~\bibnamefont
  {Akerib}}, \bibinfo {author} {\bibfnamefont {M.}~\bibnamefont {Dragowsky}},
  \bibinfo {author} {\bibfnamefont {D.}~\bibnamefont {Driscoll}}, \bibinfo
  {author} {\bibfnamefont {S.}~\bibnamefont {Kamat}}, \bibinfo {author}
  {\bibfnamefont {T.}~\bibnamefont {Perera}}, \bibinfo {author} {\bibfnamefont
  {R.}~\bibnamefont {Schnee}}, \bibinfo {author} {\bibfnamefont
  {G.}~\bibnamefont {Wang}}, \bibinfo {author} {\bibfnamefont {R.}~\bibnamefont
  {Gaitskell}}, \bibinfo {author} {\bibfnamefont {L.}~\bibnamefont
  {Bogdanova}}, and\ \bibinfo {author} {\bibfnamefont {V.}~\bibnamefont
  {Trofimov}},\ }\href {\doibase 10.1016/j.nima.2003.11.283} {\bibfield
  {journal} {\bibinfo  {journal} {Nucl. Instrum. Methods Phys. Res., Sect. A}\
  }\textbf {\bibinfo {volume} {520}},\ \bibinfo {pages} {163} (\bibinfo {year}
  {2004})}\BibitemShut {NoStop}%
\bibitem [{\citenamefont {Isaila}\ \emph {et~al.}(2012)\citenamefont {Isaila},
  \citenamefont {Ciemniak}, \citenamefont {Feilitzsch}, \citenamefont
  {G\"{u}tlein}, \citenamefont {Kemmer}, \citenamefont {Lachenmaier},
  \citenamefont {Lanfranchi}, \citenamefont {Pfister}, \citenamefont {Potzel},
  \citenamefont {Roth}, \citenamefont {Sivers}, \citenamefont {Strauss},
  \citenamefont {Westphal},\ and\ \citenamefont {Wiest}}]{Isaila2012}%
  \BibitemOpen
  \bibfield  {author} {\bibinfo {author} {\bibfnamefont {C.}~\bibnamefont
  {Isaila}}, \bibinfo {author} {\bibfnamefont {C.}~\bibnamefont {Ciemniak}},
  \bibinfo {author} {\bibfnamefont {F.}~\bibnamefont {Feilitzsch}}, \bibinfo
  {author} {\bibfnamefont {A.}~\bibnamefont {G\"{u}tlein}}, \bibinfo {author}
  {\bibfnamefont {J.}~\bibnamefont {Kemmer}}, \bibinfo {author} {\bibfnamefont
  {T.}~\bibnamefont {Lachenmaier}}, \bibinfo {author} {\bibfnamefont {J.-C.}\
  \bibnamefont {Lanfranchi}}, \bibinfo {author} {\bibfnamefont
  {S.}~\bibnamefont {Pfister}}, \bibinfo {author} {\bibfnamefont
  {W.}~\bibnamefont {Potzel}}, \bibinfo {author} {\bibfnamefont
  {S.}~\bibnamefont {Roth}}, \bibinfo {author} {\bibfnamefont {M.}~\bibnamefont
  {Sivers}}, \bibinfo {author} {\bibfnamefont {R.}~\bibnamefont {Strauss}},
  \bibinfo {author} {\bibfnamefont {W.}~\bibnamefont {Westphal}}, and\ \bibinfo
  {author} {\bibfnamefont {F.}~\bibnamefont {Wiest}},\ }\href {\doibase
  10.1016/j.physletb.2012.08.003} {\bibfield  {journal} {\bibinfo  {journal}
  {Phys. Lett. B}\ }\textbf {\bibinfo {volume} {716}},\ \bibinfo {pages} {160}
  (\bibinfo {year} {2012})}\BibitemShut {NoStop}%
\bibitem [{\citenamefont {Spooner}\ \emph {et~al.}(1992)\citenamefont
  {Spooner}, \citenamefont {Homer},\ and\ \citenamefont {Smith}}]{Spooner1992}%
  \BibitemOpen
  \bibfield  {author} {\bibinfo {author} {\bibfnamefont {N.}~\bibnamefont
  {Spooner}}, \bibinfo {author} {\bibfnamefont {G.}~\bibnamefont {Homer}}, and\
  \bibinfo {author} {\bibfnamefont {P.}~\bibnamefont {Smith}},\ }\href
  {\doibase 10.1016/0370-2693(92)90211-L} {\bibfield  {journal} {\bibinfo
  {journal} {Phys. Lett. B}\ }\textbf {\bibinfo {volume} {278}},\ \bibinfo
  {pages} {382} (\bibinfo {year} {1992})}\BibitemShut {NoStop}%
\bibitem [{\citenamefont {Basu~Thakur}(2015)}]{BasuThakur2014}%
  \BibitemOpen
  \bibfield  {author} {\bibinfo {author} {\bibfnamefont {R.}~\bibnamefont
  {Basu~Thakur}},\ }\emph {\bibinfo {title} {{The Cryogenic Dark Matter Search
  low ionization-threshold experiment}}},\ \href
  {http://lss.fnal.gov/archive/thesis/2000/fermilab-thesis-2014-33.shtml}
  {Ph.D. thesis},\ \bibinfo  {school} {University of Illinois at
  Urbana-Champaign} (\bibinfo {year} {2015})\BibitemShut {NoStop}%
\bibitem [{\citenamefont {Michael}\ \emph {et~al.}(2008)\citenamefont {Michael}
  \emph {et~al.}}]{Michael2008}%
  \BibitemOpen
  \bibfield  {author} {\bibinfo {author} {\bibfnamefont {D.}~\bibnamefont
  {Michael}}\  \emph {et~al.} (\bibinfo {collaboration} {MINOS
  Collaboration}),\ }\href {\doibase
  http://dx.doi.org/10.1016/j.nima.2008.08.003} {\bibfield  {journal} {\bibinfo
   {journal} {Nucl. Instrum. Methods Phys. Res., Sect. A}\ }\textbf {\bibinfo
  {volume} {596}},\ \bibinfo {pages} {190 } (\bibinfo {year}
  {2008})}\BibitemShut {NoStop}%
\bibitem [{\citenamefont {Hampel}\ and\ \citenamefont
  {Remsberg}(1985)}]{Hampel1985}%
  \BibitemOpen
  \bibfield  {author} {\bibinfo {author} {\bibfnamefont {W.}~\bibnamefont
  {Hampel}}\ and\ \bibinfo {author} {\bibfnamefont {L.~P.}\ \bibnamefont
  {Remsberg}},\ }\href {\doibase 10.1103/PhysRevC.31.666} {\bibfield  {journal}
  {\bibinfo  {journal} {Phys. Rev. C}\ }\textbf {\bibinfo {volume} {31}},\
  \bibinfo {pages} {666} (\bibinfo {year} {1985})}\BibitemShut {NoStop}%
\bibitem [{\citenamefont {Bearden}\ and\ \citenamefont
  {Burr}(1967)}]{Bearden1967}%
  \BibitemOpen
  \bibfield  {author} {\bibinfo {author} {\bibfnamefont {J.}~\bibnamefont
  {Bearden}}\ and\ \bibinfo {author} {\bibfnamefont {A.}~\bibnamefont {Burr}},\
  }\href {\doibase 10.1103/RevModPhys.39.125} {\bibfield  {journal} {\bibinfo
  {journal} {Rev. Mod. Phys.}\ }\textbf {\bibinfo {volume} {39}},\ \bibinfo
  {pages} {125} (\bibinfo {year} {1967})}\BibitemShut {NoStop}%
\bibitem [{\citenamefont {Sch{\"{o}}nfeld}(1998)}]{Schonfeld1998}%
  \BibitemOpen
  \bibfield  {author} {\bibinfo {author} {\bibfnamefont {E.}~\bibnamefont
  {Sch{\"{o}}nfeld}},\ }\href {\doibase 10.1016/S0969-8043(97)10073-2}
  {\bibfield  {journal} {\bibinfo  {journal} {Appl. Radiat. Isot.}\ }\textbf
  {\bibinfo {volume} {49}},\ \bibinfo {pages} {1353} (\bibinfo {year}
  {1998})}\BibitemShut {NoStop}%
\bibitem [{\citenamefont {Lindhard}\ \emph
  {et~al.}(1963{\natexlab{a}})\citenamefont {Lindhard}, \citenamefont
  {Nielsen}, \citenamefont {Scharff},\ and\ \citenamefont
  {Thomsen}}]{Lindhard1963}%
  \BibitemOpen
  \bibfield  {author} {\bibinfo {author} {\bibfnamefont {J.}~\bibnamefont
  {Lindhard}}, \bibinfo {author} {\bibfnamefont {V.}~\bibnamefont {Nielsen}},
  \bibinfo {author} {\bibfnamefont {M.}~\bibnamefont {Scharff}}, and\ \bibinfo
  {author} {\bibfnamefont {P.~V.}\ \bibnamefont {Thomsen}},\ }\href
  {http://www.sdu.dk/media/bibpdf/Bind 30-39/Bind/mfm-33-10.pdf} {\bibfield
  {journal} {\bibinfo  {journal} {Mat. Fys. Medd. Dan. Vid. Selsk.}\ }\textbf
  {\bibinfo {volume} {33}},\ \bibinfo {pages} {10} (\bibinfo {year}
  {1963}{\natexlab{a}})}\BibitemShut {NoStop}%
\bibitem [{\citenamefont {Lindhard}\ \emph
  {et~al.}(1963{\natexlab{b}})\citenamefont {Lindhard}, \citenamefont
  {Scharff},\ and\ \citenamefont {Schiott}}]{Lindhard1963a}%
  \BibitemOpen
  \bibfield  {author} {\bibinfo {author} {\bibfnamefont {J.}~\bibnamefont
  {Lindhard}}, \bibinfo {author} {\bibfnamefont {M.}~\bibnamefont {Scharff}},
  and\ \bibinfo {author} {\bibfnamefont {H.~E.}\ \bibnamefont {Schiott}},\
  }\href {http://www.sdu.dk/media/bibpdf/Bind 30-39/Bind/mfm-33-14.pdf}
  {\bibfield  {journal} {\bibinfo  {journal} {Mat. Fys. Medd. Dan. Vid.
  Selsk.}\ }\textbf {\bibinfo {volume} {33}},\ \bibinfo {pages} {14} (\bibinfo
  {year} {1963}{\natexlab{b}})}\BibitemShut {NoStop}%
\bibitem [{\citenamefont {Lindhard}\ \emph {et~al.}(1968)\citenamefont
  {Lindhard}, \citenamefont {Nielsen},\ and\ \citenamefont
  {Scharff}}]{Lindhard1968}%
  \BibitemOpen
  \bibfield  {author} {\bibinfo {author} {\bibfnamefont {J.}~\bibnamefont
  {Lindhard}}, \bibinfo {author} {\bibfnamefont {V.}~\bibnamefont {Nielsen}},
  and\ \bibinfo {author} {\bibfnamefont {M.}~\bibnamefont {Scharff}},\ }\href
  {http://www.sdu.dk/media/bibpdf/Bind 30-39/Bind/mfm-36-10.pdf} {\bibfield
  {journal} {\bibinfo  {journal} {Mat. Fys. Medd. Dan. Vid. Selsk.}\ }\textbf
  {\bibinfo {volume} {36}},\ \bibinfo {pages} {10} (\bibinfo {year}
  {1968})}\BibitemShut {NoStop}%
\bibitem [{\citenamefont {Barker}\ and\ \citenamefont
  {Mei}(2012)}]{Barker2012}%
  \BibitemOpen
  \bibfield  {author} {\bibinfo {author} {\bibfnamefont {D.}~\bibnamefont
  {Barker}}\ and\ \bibinfo {author} {\bibfnamefont {D.-M.}\ \bibnamefont
  {Mei}},\ }\href {\doibase 10.1016/j.astropartphys.2012.08.006} {\bibfield
  {journal} {\bibinfo  {journal} {Astropart. Phys.}\ }\textbf {\bibinfo
  {volume} {38}},\ \bibinfo {pages} {1} (\bibinfo {year} {2012})}\BibitemShut
  {NoStop}%
\bibitem [{\citenamefont {Jones}\ and\ \citenamefont
  {Kraner}(1975)}]{Jones1975}%
  \BibitemOpen
  \bibfield  {author} {\bibinfo {author} {\bibfnamefont {K.~W.}\ \bibnamefont
  {Jones}}\ and\ \bibinfo {author} {\bibfnamefont {H.~W.}\ \bibnamefont
  {Kraner}},\ }\href {\doibase 10.1103/PhysRevA.11.1347} {\bibfield  {journal}
  {\bibinfo  {journal} {Phys. Rev. A}\ }\textbf {\bibinfo {volume} {11}},\
  \bibinfo {pages} {1347} (\bibinfo {year} {1975})}\BibitemShut {NoStop}%
\bibitem [{\citenamefont {Barbeau}\ \emph {et~al.}(2007)\citenamefont
  {Barbeau}, \citenamefont {Collar},\ and\ \citenamefont
  {Tench}}]{Barbeau2007}%
  \BibitemOpen
  \bibfield  {author} {\bibinfo {author} {\bibfnamefont {P.~S.}\ \bibnamefont
  {Barbeau}}, \bibinfo {author} {\bibfnamefont {J.~I.}\ \bibnamefont {Collar}},
  and\ \bibinfo {author} {\bibfnamefont {O.}~\bibnamefont {Tench}},\ }\href
  {\doibase 10.1088/1475-7516/2007/09/009} {\bibfield  {journal} {\bibinfo
  {journal} {J. Cosmol. Astropart. Phys.}\ }\bibinfo {volume} {09} (\bibinfo
  {year} {2007})\ \bibinfo {pages} {009}}\BibitemShut {NoStop}%
\bibitem [{\citenamefont {Yellin}(2002)}]{Yellin2002}%
  \BibitemOpen
  \bibfield  {author} {\bibinfo {author} {\bibfnamefont {S.}~\bibnamefont
  {Yellin}},\ }\href {\doibase 10.1103/PhysRevD.66.032005} {\bibfield
  {journal} {\bibinfo  {journal} {Phys. Rev. D}\ }\textbf {\bibinfo {volume}
  {66}},\ \bibinfo {pages} {032005} (\bibinfo {year} {2002})}\BibitemShut
  {NoStop}%
\bibitem [{\citenamefont {Yellin}(2007)}]{Yellin2007}%
  \BibitemOpen
  \bibfield  {author} {\bibinfo {author} {\bibfnamefont {S.}~\bibnamefont
  {Yellin}},\ }\href@noop {} {}\Eprint {http://arxiv.org/abs/0709.2701}
  {arXiv:0709.2701 [physics.data-an]} \BibitemShut {NoStop}%
\bibitem [{\citenamefont {Lewin}\ and\ \citenamefont
  {Smith}(1996)}]{Lewin1996}%
  \BibitemOpen
  \bibfield  {author} {\bibinfo {author} {\bibfnamefont {J.}~\bibnamefont
  {Lewin}}\ and\ \bibinfo {author} {\bibfnamefont {P.}~\bibnamefont {Smith}},\
  }\href {\doibase 10.1016/S0927-6505(96)00047-3} {\bibfield  {journal}
  {\bibinfo  {journal} {Astropart. Phys.}\ }\textbf {\bibinfo {volume} {6}},\
  \bibinfo {pages} {87} (\bibinfo {year} {1996})}\BibitemShut {NoStop}%
\end{thebibliography}%

\end{document}